\documentclass[a4paper,twocolumn,english,aps,prb,superscriptaddress,showpacs,showkeys]{revtex4} \usepackage[T1]{fontenc}
\usepackage[latin9]{inputenc}
\usepackage{amsmath}
\usepackage{graphicx}
\usepackage[sort&compress]{natbib}
\usepackage{hyperref}

\hypersetup{
    colorlinks=true,        % false: boxed links; true: colored links
    linkcolor=blue,          % color of internal links
    citecolor=blue,        % color of links to bibliography
    filecolor=blue,      % color of file links
    urlcolor=black           % color of external links
}

\makeatletter
\@ifundefined{textcolor}{}
{%
 \definecolor{BLACK}{gray}{0}
 \definecolor{WHITE}{gray}{1}
 \definecolor{RED}{rgb}{1,0,0}
 \definecolor{GREEN}{rgb}{0,1,0}
 \definecolor{BLUE}{rgb}{0,0,1}
 \definecolor{CYAN}{cmyk}{1,0,0,0}
 \definecolor{MAGENTA}{cmyk}{0,1,0,0}
 \definecolor{YELLOW}{cmyk}{0,0,1,0}
 }

\usepackage{subfigure}
\@ifundefined{definecolor}
 {\@ifundefined{definecolor}
 {\usepackage{color}}{}
}{}

\makeatother
\usepackage{babel}
\makeatother
\usepackage{babel}

\begin{document}
\date{\today}

\title{Modelling impurity-assisted chain creation in noble-metal break junctions}

\author{Solange Di Napoli}
\affiliation{Departamento de F\'{\i}sica de la Materia Condensada, CAC-CNEA, Avenida
General Paz 1499, (1650) San Mart\'{\i}n, Pcia. de Buenos Aires, Argentina}
\affiliation{Consejo Nacional de Investigaciones Cient\'{\i}ficas y Técnicas,
CONICET, Buenos Aires, Argentina}

\author{Alexander Thiess}
\author{Stefan Bl\"ugel}
\author{Yuriy Mokrousov}
\affiliation{Peter Gr\"unber Institut and Institute for Advanced Simulation, Forschungszentrum J\"ulich
and JARA, D-52425 J\"ulich, Germany}

\begin{abstract}
In this work we present the generalization of the model for chain formation in 
break-junctions, introduced by Thiess {\it et al.}~[\onlinecite{Alex08}], to 
zigzag transition-metal chains with $s$ and $p$ impurities. We apply this extended model to
study the producibility trends for noble-metal chains with impurities, often present
in break junction experiments, namely, Cu, Ag and Au chains with H, C, O and N adatoms.
Providing the material-specific parameters for our model from systematic full-potential 
linearized augmented plane-wave first-principles calculations, we find that the presence 
of such impurities crucially affects the binding properties of the noble-metal chains. 
We reveal that both, the impurity-induced bond strengthening and the formation of zigzag bonds, can lead to
a significantly enhanced probability for chain formation in break junctions.
\end{abstract}

\maketitle

\section{Introduction}

Quantum transport in nanoscale and atomic-scale objects has recently 
become one of the most researched areas in modern physics, owing to
vast possible applications in technology. On the most exciting side, 
electronic properties of one-dimensional (1D) systems such as atomic
chains and molecular wires often lead to fascinating transport properties,
which were predicted to occur theoretically and could be observed
experimentally. After years of research, the material versatility of the
class of 1D systems still remains rather poor, however. This can be mainly
attributed to the fact that to produce and sustain a free-standing, or 
even deposited chain proved to be rather difficult in experiment. 
One of the most promising experimental
techniques developed over the last decade, which allows to produce 1D 
structures and measure their transport properties, lies in investigating
the mechanically controllable break-junctions (MCBJ).\cite{Ruitenbeek01}   

In a typical break junction experiment, two pieces of material being initially
in contact are pulled apart while the conductance of the system is measured. 
During the process of pulling, several atoms may be extracted from the leads, 
forming a free-standing finite 1D chain.\cite{Ruitenbeek01} With this technique
it has been possible to create monoatomic chains with the length of 5 to 10 atoms
in Ir,~\cite{Ryu06} Pt~\cite{Ruitenbeek01} and Au~\cite{Ohnishi98,Ruitenbeek98,Rubio01}
break junctions. The ability to create similarly long chains in break junctions
of other metals has been greatly hindered so far.~E.g.~it has been demonstrated
that the tendency for chain creation in lighter $4d$ and $5d$ transition-metals (TMs)
is significantly lower than in $5d$ TMs. This trend was later explained based
on the relativistic enhancement of mass in heavier elements with corresponding analogy
to the reconstruction of the low-index surfaces of the $5d$ TMs.

In the past years it became evident that it is possible to strengthen the bonds 
in a suspended chain and achieve higher degree of chain's producibility by 
adding external absorbates during the process of chain formation. This opens
a new possibility for achieving desired versatility of produced in MCBJ 
experiments one-dimensional systems with different properties.  
%Moreover, indirect evidence has recently been obtained for oxygen atoms inserted
%into metallic linear atomic chains (LAC)~\cite{Thijssen06}.
In a key experiment,\cite{Thijssen06} Thijssen \textit{et al.} observed such
an enhanced tendency towards chain elongation for Au, Ag and Cu break junctions
with oxygen present in the atmosphere. Moreover, the effect of other impurities,
such as H, B, C, N and S, on the creation of Au chains has been considered,
while stable and strong bonding between the Au and impurity atoms has been
demonstrated.\cite{Novaes03,Barnett04,Ugarte02}
Physically, such a situation can be explained based on a well-known observation 
that low-coordinated noble metal (NM) atoms are chemically more reactive than in the 
bulk and thus their chains are expected to be even more reactive 
than nanoparticles.\cite{Novaes06,Bahn02}

Experimental advances in producing longer chains with the aid of 
impurities of various physical and chemical nature call for an appropriate 
theoretical description. Taking into account the complexity of MCBJ experiments,
the aim of the proper chain formation theory lies mainly in guiding future 
successful experiments and extending the material base for further studies. In this respect the
predictive power of density functional theory (DFT) first-principles, or, 
{\it ab initio}, techniques is particularly valuable. Treating the complicated
process of chain's stretching, elongation and eventual breaking upon
pulling the electrodes apart in an atmosphere contaminated with various
elements, is an extremely challenging task even for the fastest first-principles
methods. The main obstacle here is the necessity to describe simultaneous 
restructuring of hundreds of atoms in real time. Due to computational limitations, 
{\it ab initio} studies of impurity-assisted chain formation so far focused mainly 
on stability, mechanical and electronic properties of clean and contaminated {\it infinite} chains 
of transition metals.\cite{Ugarte04,Artacho99,Alonso05,Calzolari04} This situation calls for a development of an effective model 
which is able to utilize the results of first-principles calculations of simplified
systems for the purpose of describing the chain formation on a level, which allows to
extend the material base for future MCBJ experiments with a reasonable computational effort. 

Recently, Thiess and co-authors introduced a model ('CF model', or, 'chain formation model' in the 
following) able to provide a probability
of chain's elongation in a break junction using only a limited number of parameters
which can be extracted from first-principles calculations of a perfect infinite
chain and idealized leads. Using this model, the authors of ref.~\cite{Alex08}
analyzed the chemical trend of chain creation in break junctions of $3d$, $4d$ 
and $5d$ transition metals, arriving at a qulitative agreement with experiments 
and unravelling the physics behind the observed trends.\cite{Alex08} An attractive feature
of this approach lies in the ability to visualize the process of chain formation
by referring to phase diagrams for probability of elongation as a function of 
chain's stretching and the number of chain's atoms. Additionally, the obtained
chemical trends can be understood based on a small number of element-specific
parameters. Using this model, the tendency for chain elongation in linear Au-O chains
has been predicted to be stronger than in pure Au chains, in accordance to 
experiments.\cite{Alex08} 

In this work, we systematically apply the CF model for a detailed study of
the trends in formation of Cu, Ag and Au chains in pure break junctions as well 
as Cu, Ag and Au break junctions with atmosphere contaminated by H, O, N, and C 
impurities. Additionally, we extend the CF model to the case of geometrically 
more complex zigzag arrangement of the atoms in the chain. In particular, we 
find that some of the chains exhibit a finite probability for elongation under
significant tension in the linear regime as well as for the zigzag arrangement.
While the probability of the chain's elongation in the zigzag regime is rather
small, it has an advantage that it is more robust with respect to breaking. Overall,
we predict that incorporation of all considered impurities, even hydrogen, into 
the noble metal (NM) chains results in an increased probability for chain elongation when 
compared to the pure chains of NM elements. We also observe that although NM chains
with $s$-type H impurities display wide regions of finite probability for elongation as
a function of strain, this probability is much smaller than that for NM chains
with $p$-type impurities. Remarkably, our calculations predict a noticeably    
higher tendency towards chain elongation in Cu and Au contaminated chains, than
in Ag chains. 

The paper is organized as follows. In the next section we provide the details
of our DFT first-prinicples calculations. In Section~\ref{single-atom} we analyze
the energetics and structural properties of linear, zigzag and dimerized single-atom 
NM chains and NM chains with O, H, C, and N impurities. After this in Section~\ref{P-model}
we present the definition and the details of the model we use to estimate the tendency
for chain elongation, as well as its generalization to the case of the 
zigzag geometry. Finally, in Section~\ref{Prod-M-X} we analyze the process
of chain formation in break junctions of NMs, pure and with impurities, and compare 
the predicted trends. A summary and conclusions are given in Section~\ref{conclusions}.

\section{Computational details}

\label{comp-det} 

\begin{figure}[t!]
\begin{centering}
\includegraphics[width=8.5cm]{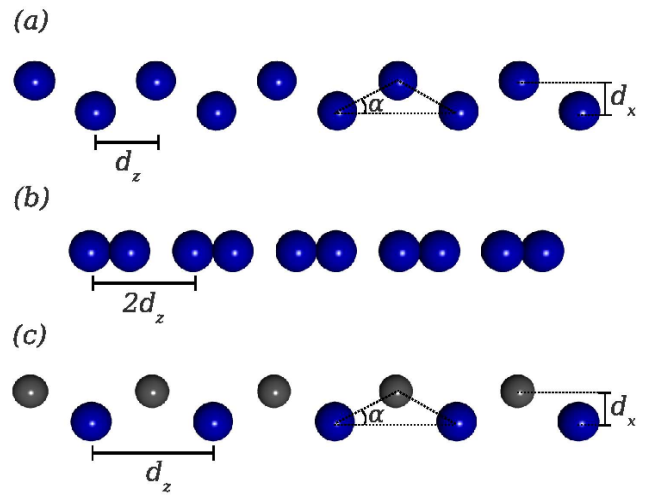}
\par\end{centering}
\caption{(Color online) Schematic structure of (a) single-atom planar zigzag
chains, (b) single-atom dimerized chains and (c) planar zigzag chains
with impurities. The noble atoms are represented with large blue spheres,
while smaller grey spheres stand for the impurity atoms.}
\label{definitions}
\end{figure}

For our density functional theory first-principles calculations we employed 
the full-potential linearized augmented plane-wave (FLAPW) method
for 1D systems,\cite{Yura1D} as implemented
in the {\tt FLEUR} code~\cite{FLEUR} within the generalized gradient 
approximation (GGA) to the exchange-correlation potential. The FLAPW 
basis functions were expanded up to $k_{\rm max}$ of 4.0~bohr$^{-1}$ and
we used 32 $k$ points in one half of the 1D Brillouin zone for our 
self-consistent calculations. The diameter of the cylindrical boundary 
between the interstitial and vacuum regions, $D_{\rm vac}$, and the 
fictitous in-plane lattice constant, $\tilde{D}$, were set to 8.3~bohr 
and 9.9~bohr, respectively. The muffin-tin radii have been set to 
1.90~bohr for Au (with local orbitals for the $5s$ and $5p$ states),
2.0~bohr for Ag and Cu, and 1.0~bohr for the impurities. These values 
were chosen to guarantee the required accuracy as well as to enable a 
comparison between the total energies of the zigzag chains with a wide 
range of the angle $\alpha$, see Fig.~\ref{definitions}. We have set the 
coordinate system such that the chain axis is aligned along the $z$ axis, 
and considered a two-atoms unit cell to allow for zigzag and dimer 
arrangements. In Fig.~\ref{definitions} a sketch of the infinite planar 
zigzag and dimerized atomic chains is shown, where the definition of 
$d_{z}$ and $d_{x}$ distances, as well as the zigzag angle $\alpha$
are given. 
For our calculations we neglected the effect of the spin-orbit 
interaction, since its effect on the energetics of the chain formation was
shown to be negligible.\cite{Alex09,Alex10}

\section{Energetics}
\label{single-atom}

One of the purposes of our work is to understand the role of the
geometric structure in chain formation taking
into account energetics considerations.  From this starting
point we analyze the particular role of environmental impurities in the
chain formation process. In order to achieve this goal, at first, we perform 
systematic total energy calculations as a function of the both interatomic
parameters, $d_{z}$ and $d_{x}$ (c.f.~Fig.~\ref{definitions}). 
For each $d_{z}$ we performed a total energy calculation for at least five 
different $d_{x}$ values.
On the basis of these calculations we interpolated the two-dimensional energy 
profile in the $d_z$-$d_x$ plane, from which we extract the effective minimum
energy curves as a function of $d_z$, $\varepsilon(d_z)=\min_{d_x}\varepsilon(d_z,d_x)$.

\begin{figure}[t!]
\begin{centering}
\includegraphics[width=8.5cm]{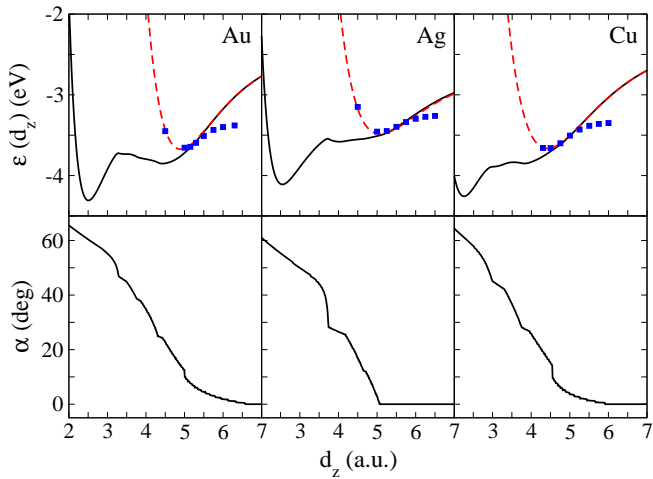}
\par\end{centering}
\caption{(Color online) Upper panel: The minimal total energy $\varepsilon$ (per atom) as a function of
$d_{z}$ for Au (left), Ag (middle) and Cu (right). The dashed red lines correspond to the
total energy of the linear infinite single-atom chains, while the filled squares are the
total energies of the dimerized chains. Lower panel: The evolution of the corresponding
zigzag $\alpha$-angles.}
\label{AuAgCu}
\end{figure}

Our results for the noble metal Cu, Ag and Au single-atom chains are summarized 
in Fig.~\ref{AuAgCu}, where the total energy and the zigzag $\alpha$-angles are 
plotted as a function of $d_{z}$. The corresponding energies of the single-atom 
infinite linear and dimerized chains are also presented for comparison. As can be 
seen from this figure, in all three cases the zigzag structures with large values
of the $\alpha$-angle are more stable than the linear configurations, in agreement 
with previous studies.~\cite{Ferrer07,Sanchez01,Cakir11}
All energy profiles display a characteristic two-well structure, where the first 
minimum correspond to the $\alpha$-angles in the vicinity of 60$^{\circ}$.
In this configuration, each atom interacts strongly not only with its "chain" first
nearest neighbors along the $z$-axis, but also with the second nearest neighbors.
The second minimum in the total energy curves, which is clearly pronounced in 
particular for Au and Cu, occurs
for the values of $\alpha$ at around $25^{\circ}$, corresponding to
a zigzag structure with each atom interacting mainly with its two first nearest neighbors.
While the zigzag angle falls off rapidly to zero for $d_z$ above 5.0 bohr for all chains, this
decay is more abrupt for the case of Ag chains.
As far as the dimerization is concerned, we find that the chains tend to dimerize when 
elongated above the $d_z$ of $\approx 5.5$ bohr, with a modest gain in energy as compared
to the linear configuration for $5.5 <d_z< 6.5$ bohr. As we shall see in the following, when
the interatomic distance lies in this interval, the chains are already very prone to breaking, thus,
we do not consider the effect of dimerization on the chain formation considered in the following
sections.

\begin{figure}[t!]
\begin{centering}
\includegraphics[width=8.7cm]{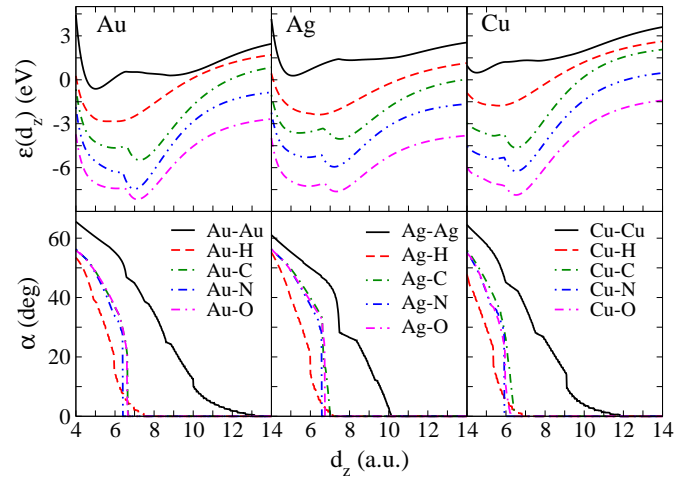}
\par\end{centering}
\caption{(Color online) Upper panel: The minimal total energy (per unit cell) as a function of $d_z$
for Au (left), Ag (middle) and Cu (right). For better comparison, the energy curves were rigidly shifted with respect to each other.
Lower panel: The evolution of the corresponding zigzag $\alpha$-angles.}
\label{X-NM}
\end{figure}

Next, we follow the same procedure for the case with an impurity in the two-atom unit cell, 
as shown schematically in Fig.~\ref{definitions}(c). In Fig.~\ref{X-NM} we present the evolution 
of the minimal total energy and corresponding zigzag angle with $d_z$ in Cu, Ag and
Au chains with H, C, N, and O impurities (NM-X chains). Note, that the energy profiles
are shifted in this figure with respect to each other along the $y$ axis for better visibility.
From Fig.~\ref{X-NM} we observe that the behaviour of the total energy is different as a function of $d_z$
for $s$- (H) and $p$-impurities (C, N, O). In the latter case the global minimum in energy occurs
for the linear configuration ($\alpha=0^{\circ}$) and the second local energy minimum can be seen 
also for the zigzag planar geometry with comparatively large angles $\alpha$ in the range of $30-40^{\circ}$. 
This is in contrast to 
the pure NM chains, for which two energy minima are also present, yet they both correspond to zigzag 
geometry with profound energetic preference to higher $\alpha$-angles, see Fig.~\ref{AuAgCu}.
On the other hand, for NM-H chains the zigzag structure exhibits the global and only energy minimum, although 
the values of the zigzag angle are rather small, namely, $\alpha\approx 12^\circ$ (Au-H), $\alpha \approx 6^\circ$ (Ag-H)
and $\alpha \approx 10^\circ$ (Cu-H). We can therefore conclude that the presence of impurities in NM break
junctions leads to an effective straightening of the bonds in the chain with energetically favored configurations which
are linear or exhibiting small zigzag angles.

A common feature for all Ag-X chains, which can be seen in Fig.~\ref{X-NM}, is the
smaller slope of the energy curves as a function of $d_z$, when compared to corresponding
Cu-X and Au-X cases. Such "softening" of the $\varepsilon(d_z)$ profile can be seen
already for pure Ag chains, when compared to Au and Cu chains, especially in the vicinity of
the global zigzag minimum, Fig.~\ref{AuAgCu}. The reason behind this phenomenon lies in the
fact that the $d$ states of Ag are situated much lower in energy than the $s$-states, which 
solely determine the character of the electronic states in the vicinity of the Fermi energy ($E_F$).
In case of Au and Cu, the $d$ states are positioned higher in energy and overlap with the $s$ states
at the $E_F$, providing additional channel for hybridization and bonding between the atoms.
Correspondingly, the bonding in the chains of Ag atoms is somewhat weaker than in those made 
of Au and Cu. Such weakening of bonding and softening of the energy profile for Ag in
comparison to Au and Cu was also observed for linear infinite chains and in 
bulk,~c.f.~[\onlinecite{Zarechnaya}] and citations therein. 

\begin{figure}
\noindent \begin{centering}
\includegraphics[width=8.5cm]{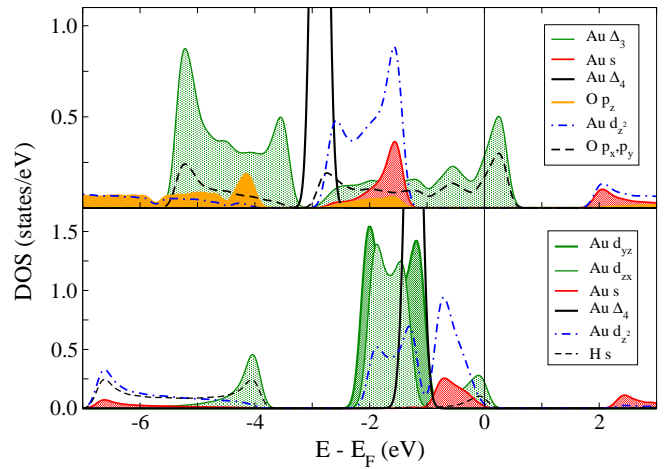}
\par\end{centering}
\caption{(Color online) Atomically-resolved density of states for Au-O chain with $d_z=7.25$ bohr and 
$\alpha=0^\circ$ (top), and Au-H chain with $d_z=6.0$ bohr and $\alpha\approx 18^\circ$ (bottom). The densities
of states are additionally decomposed into the contributions coming from $s$, $p$ and $d$ states of
different symmetry.}
\label{dos-AuHvsAuO}
\end{figure}

In order to understand the difference in structural properties of NM chains with $s$- and $p$- impurities, we 
consider the density of states (DOS) of Au-O ($d_z=7.25$ bohr, $\alpha=0^{\circ}$) and Au-H ($d_z=6.0$ bohr, 
$\alpha\approx 18^\circ$) chains in close to equilibrium cofigurations, presented in Fig.~\ref{dos-AuHvsAuO}. 
In this plot, the DOS is additionally
decomposed into atomic (Au or X) and orbital $s$, $p$ and $d$ contributions. By looking at the DOS of the Au-O chain first,
we observe that for the linear arrangement of the atoms in the chain the bonding of the atoms of Au and O is
the strongest owing to the pronounced directionality of the oxygen's valence $p$-orbitals. In this case the
overlap between the $s$ and $d_{z^2}$ orbitals of Au and $p_z$ orbital of O along the chain axis, as well as 
$p_x$ and $p_y$ oxygen orbitals with Au $d$ states of the $\Delta_3$ symmetry ($d_{zx}$ and $d_{yz}$ orbitals)
is the largest, which can be seen from the DOS (upper panel in Fig.~\ref{dos-AuHvsAuO}). This leads to a 
pronounced minimum in the total energy for the linear regime. On the other hand, the hydrogen $s$ orbital 
lacks directionality of the $p$ orbitals of oxygen atoms. In the linear case, brought into contact with
Au atoms, the H $s$-state can hybridize only with the corresponding Au states of the same $\Delta_1$ symmetry,
namely, $6s$ and $5d_z$ states, while the states of the $\Delta_3$ and $\Delta_4$ ($d_{xy}$ and $d_{x^2-y^2}$)
symmetry would remain non-bonding. Upon transition to a zigzag configuration, the cylindrical symmetry of the chain
is broken and the hybridyzation between the hydrogen $s$ and Au $\Delta_3$ orbitals occurs since it is not anymore 
forbidden by symmetry, as can be seen from the DOS in lower panel of Fig.~\ref{dos-AuHvsAuO}. This additional
channel for bonding in the chain energetically favors the zigzag arrangement of H and Au atoms, although the
magnitude of the zigzag angle is limited by the small radius of the $1s$ hydrogen orbital. Noticeably, 
in both cases the $\Delta_4$ Au orbitals, which lie in the plane perpendicular to the chain axis, remain
non-bonding, Fig.~\ref{dos-AuHvsAuO}.

\section{Model of chain formation}
\label{P-model}

During the process of stretching and chain formation in a MCBJ experiment, the atoms 
are extracted from the leads upon pulling them apart, and form a chain consisting of 
several atoms. Owing to the complex relaxations of hundreds of atoms during this
process, its rigorous description from the first-principles is an extremely challenging
task. Moreover, the exact structure of the electrodes is normally unknown,
which introduces a great deal of uncertainty in the initial configuration of the system. 
In the past, several ways to overcome these obstacles were taken, among which 
the most successful ones lie in employing the molecular dynamics technique 
to tackle the chain formation process. \cite{Ugarte04,daSilva08,Hasmy08,daSilva07,Soler07}In this work we choose a different path, and 
extend a simple model of chain formation proposed by Thiess and co-workers,
\cite{Alex08} which is based on general total energy arguments. The applicability
of this model to description of the complicated MCBJ chain fomation process has
been demonstrated for transition metals.\cite{Alex08,Alex09}

According to the CF model, we divide the system into two regions: the leads and the
suspended chain. The electronic structure of the two parts is considered separately
from {\it ab initio}, thus neglecting their mutual influence. This approximation becomes
increasingly better when going to the limit of longer chains, which are in the focus of
our study. The process of the chain formation is monitored by following the change in the 
total energy of an atom that is transferred from the lead into the chain upon chain 
elongation. To estimate this total energy change we assume that upon the transfer of 
the atom from the lead into the chain the only parameter which is changed in the system 
is the interatomic distance $d$ between the atoms in the chain, while the length of the chain
$L$ and the structure of the 
leads remain the same. This is the so-called rapid reformation approximation (RRA).
\cite{Rubio01,Bahn01,daSilva01,Alex08} Moreover, we suppose that the transfer of a 
lead atom into the wire happens instantaneously at a certain time $t_i$, when the 
energy of the transfered atom and its coordinates change suddenly from their values 
inside the lead to the corresponding values inside the chain. We also assume that
between two consequent elongation events at times $t_i$ and $t_{i+1}$ the pulling
of the electrodes apart results only in the smooth change of the chain's interatomic
distance $d$.

\begin{figure}[ht!]
\noindent \begin{centering}
\subfigure[]{\includegraphics[width=7.5cm]{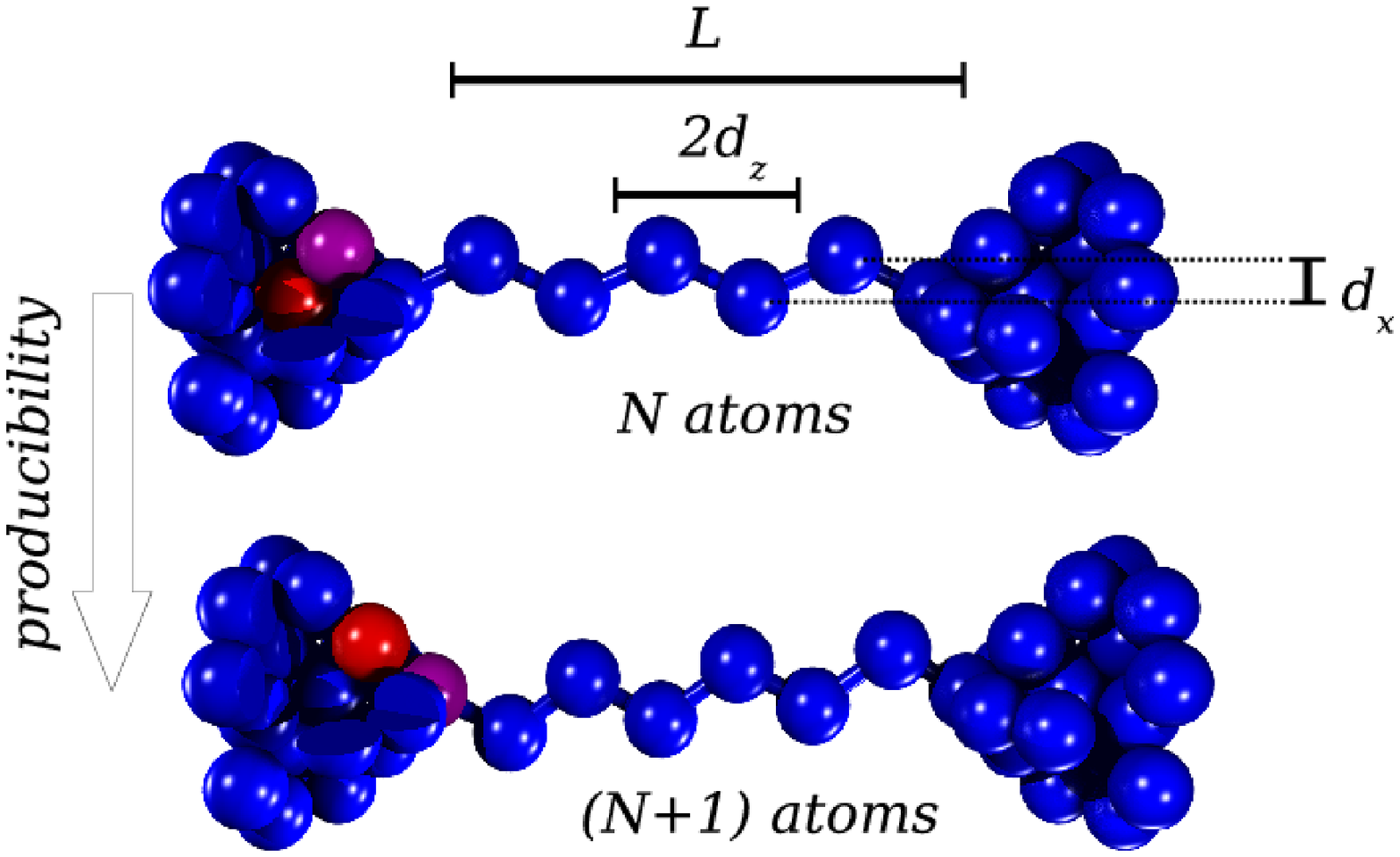}} 
\subfigure[]{\includegraphics[width=7.5cm]{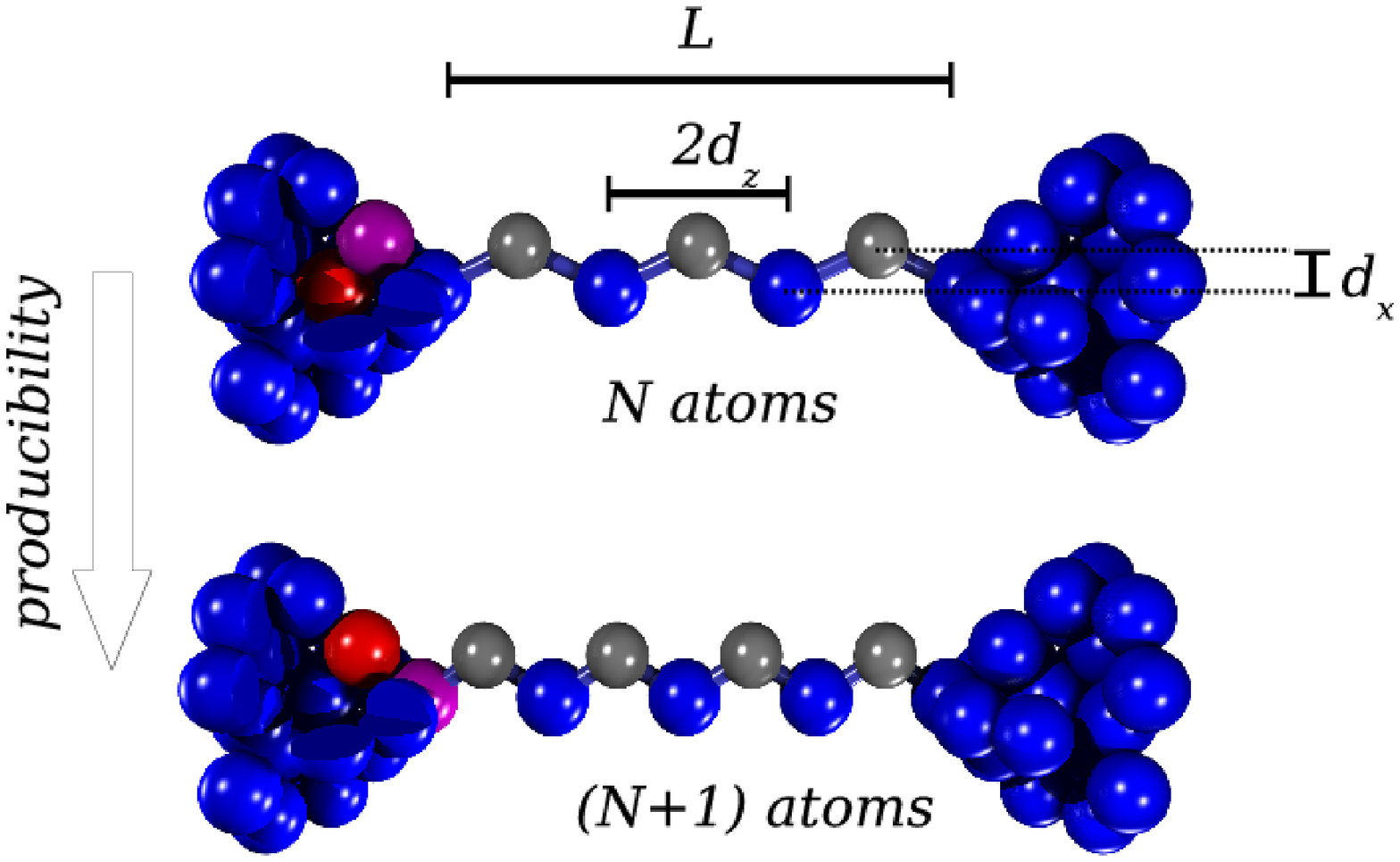}}
\end{centering}
\caption{Sketch of the producibility (P) criterion in suspended single-atom chains (a) and in suspended chains with impurities (b).}
\label{single} 
\end{figure}

In Fig.~\ref{single} we show a sketch of the system under consideration, where the two
figures for each of the situations correspond to the initial and final states of the process
of the chain elongation. 
To analyze the producibility of the chain we compare the total energies of the initial and 
the final configurations in the process of pulling one lead atom into the chain, 
where $L$ is the distance between the leads. A successful chain elongation will
be translated into the following total energy relation, the criterion for producibility 
(P-criterion):
\begin{equation}
(N+1) \epsilon (d) \ge \Delta E_{lead} + (N+2) \epsilon (\tilde{d})
\label{P}
\end{equation}
where $d=L/(N+1)$ and $\tilde{d}=L/(N+2)$ are the chain interatomic distances before 
and after the elongation, respectively, and $N+1$ is the number of bonds in a chain of 
$N$ atoms. The binding energy of the suspended chain $\epsilon(d)$ is defined from
the binding energy of the infinite wire with interatomic distance $d$, 
$E_W(d)=E_W(d_0)+\epsilon(d)$ (we always have two atoms in the unit cell), relative to the 
wire's cohesive energy at equilibrium interatomic distance $d_0$, $E_W(d_0)$, which
implies $\epsilon(d)\textgreater 0$. 

The chain formation energy in eq.~(\ref{P}) is given by 
$\Delta E_{lead}=E_W(d_0)-E_{lead}$, where $E_{lead}$ denotes the cohesive energy of 
an atom inside the lead. To calculate $E_{lead}$ we perform
a bulk calculation of the total energy for each chemical element at its corresponding 
equilibrium lattice constant, $E_{bulk}$, and a bulk 
calculation of the total energy of the isolated atom, $E_\infty$, 
to get $\Delta E_{bulk}= E_\infty - E_{bulk}$.
Taking into account the surface energies calculated by Skriver and Rosengaard,\cite{Skriver92} 
$\Delta E_{surf}=E_{surf}-E_{bulk}$,
we defined $E_{lead}=\Delta E_{bulk}-\Delta E_{surf}=E_\infty - E_{surf}$, where $E_{surf}$
is calculated analogously to $E_{bulk}$ for an atom at the surface of a certain crystallographic
orientation. At finite temperatures, $\Delta E_{lead}$
corresponds to the difference between the chemical potentials of the lead and  of the infinite
wire. Typically, $\Delta E_{lead} \textgreater 0$. If the P-criterion (eq.~(\ref{P})) 
is satisfied the system tends to increase the number of atoms in the chain by one.
One can rewrite the P-criterion in the following way:
\begin{equation}
E^*(d,N)=(N+1) \epsilon (d) - \Delta E_{lead} + (N+2) \epsilon (\tilde{d})
\label{E*}
\end{equation}
where $E^*(d,N)$ is a function of the interatomic distance and the number of atoms in the chain. If 
$E^*(d,N) \textgreater 0$ the P-criterion is satisfied and the chain can increase by one atom
for that particular pair of parameters $(d,N)$. 

To apply the P-criterion to suspended chains it is useful to fit the chain binding energy 
$\epsilon (d)$ with an analytical function. As the phase space of possible structural 
arrangements of the wires is two-dimensional, we calculated the binding energy for 
several $d_x$ and $d_z$ distances in the wire (see Figure 1), and fitted each energy curve 
with the analytical Morse potential,\cite{Alex08} interpolating the minimum energy curve from 
these data for all considered wires, as explained in Section~\ref{single-atom}. Thus, the 
binding energies participating in eqs.~(\ref{P}) and (\ref{E*}) are the energy curves shown in 
Figures \ref{AuAgCu} and \ref{X-NM}.

We define the {\it string tension}, or {\it force} $F(d)$ as the slope of the binding energy with 
respect to the distance between the atoms along the $z$-axis $d=d_z$ (see Fig.~\ref{single}):
\begin{equation}
F(d)=F(d_z)=\frac{\delta \epsilon(d_z)}{\delta d_z}
\label{string-tension}
\end{equation}     
From eqs.~(\ref{P}) and (\ref{string-tension}) we can deduce that there are 
two main parameters
which compete and which determine whether a chain can be increased by one atom.~\cite{Alex08} These
are (i) the energy cost of bringing an atom from the lead into the chain, $\Delta E_{lead}$, 
and  (ii) the {\it break force} $F_0$, which is the maximum of the string tension occuring 
at the so-called {\it inflection point} $\hat d$, that is, $F_0=F(\hat d)$. The inflection point 
gives an estimate of how far an ideal infinite chain can be stretched until it breaks by 
the maximum break force due to long-wavelength perturbations.\cite{Alex08}
Physically, a successful chain elongation event will occur when the following happens: 
as the chain is stretched, the energy of the system increases up to the point where a lead atom overcomes the chain formation barrier, $\Delta E_{lead}$, and enters the chain. This
relaxes the distance in the chain from $d$ to $\tilde d$ and lowers the total energy of an atom in the wire, $E_W(\tilde d) \textless E_W(d)$. The larger the slope of the total
energy $E_W(d)$, the more energy can be gained by relaxing the chain from a distance $d$ to $\tilde d$. Therefore, large values of $F_0$ and small energy barriers $\Delta E_{lead}$
will favor chain elongation.

When the number of atoms in the chain $N$ is very large, we can safely assume 
$\epsilon(d)-\epsilon(\tilde d) \approx \frac{\delta\epsilon(d)}{\delta d}\cdot\delta d 
= F(d)\cdot\delta d$, where $\delta d=d-\tilde d\approx \frac{L}{N^2}$,
and the P-criterion reads:
\begin{equation}
F(d) \ge \frac{\Delta E_{lead}+\epsilon(d)}{d}=f(d),
\label{P2}
\end{equation}
where $f(d)$ is the {\it generalized string tension}, or the work done in drawing the chain with interatomic distance $d$ out of the lead per unit chain length. This equation provides the
interval of distances in which the suspended chain can be producible. The analysis of the
producibility of the long wires and chains in terms of the generalized string tension has been
successfully performed in the past.\cite{Alex08,Alex09,Kondo00,Tosatti01} 
It is easy to demonstrate that the lower and upper boundaries of the interval of distances where
the chain is producible and eq.~(\ref{P2}) is satisfied, 
correspond to the local minimum and maximum of $f(d)$, respectively.

\section{Impurity-assisted chain formation in break junctions}
\label{Prod-M-X}

In this section we address the question of the impact of light $s$ and $p$ impurities 
H, N, C, and O on the producibility of noble-metal chains in a MCBJ. A key towards 
understanding the impurity assisted formation process is the additional degree of 
freedom of lateral relaxation in zigzag bonds, which can be incorporated into the 
criterion for producibility formulated in the previous section.

Before analyzing the results of our extended model, it is important to discuss crucial 
energy scales determining the elongation process. While the extraction of atoms out of the leads 
can be modeled in line with previous studies,\cite{Alex08,Alex09} we have to consider additionally
the supplementing of the pure chain with the impurities.
Here, we assume that the light impurities H, N, C, and O are present in the atmosphere and
are not bound at the surface. This assumption is well justified considering that the
concentration of such impurities is controlled by the partial pressure in experiments.
Furthermore, {\it a priori}, the disociation energy of impurity molecules potentially acts as an 
additional barrier in the energetic balance given by relation~\eqref{P}. 
Here we refer to the results by Bahn \textit{et al.}~\cite{Bahn02} who reported the chemisorption 
energy of Au chains on the order of $-$1.5 eV per O atom pair with respect to the energy of the 
free O$_2$ molecule. This finding of highly preferential binding of impurity atoms to the chain 
allows for two important simplifications.
Firstly, impurities have a strong tendency to cover all assisted bonds of the chain atoms. Hence, 
the resulting structure is well described by a chain of alternating Au and O atoms. 
Secondly, we assume that the dissociation of impurity molecules and absorption to
the metal bonds happens instantaneously after an additional metal atom has joined the 
chain. Thereby, we assume that the dissociation process happens independently of the chain formation
and we do not account for it while considering the energetics of the chain formation. This leaves us 
with energy barriers introduced as $\Delta E_{\rm lead}$ in section~\ref{P-model},
accounting for 0.86~eV, 0.75~eV and 1.16~eV for Au, Ag and Cu, respectively.

\begin{figure}
\begin{center}
\includegraphics[width=0.49\textwidth]{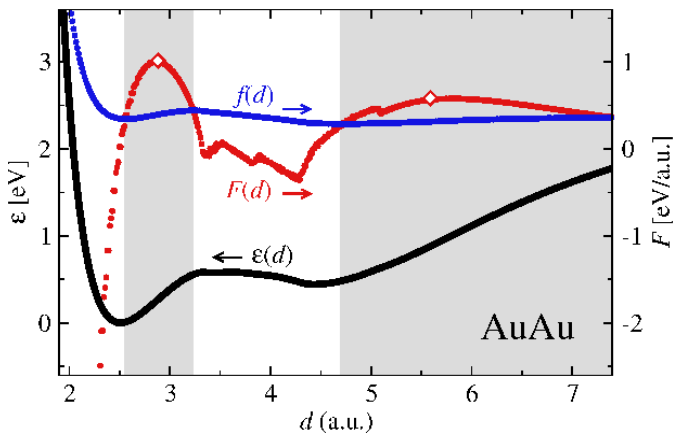}
\end{center}
\caption{(Color online). Binding energy $\varepsilon(d)$, force $F(d)$ and generalized string tension 
$f(d)$ for a pure zigzag Au chain. Both local maxima of $F(d)$ are marked with open diamonds and will 
be refered to as break force $F_0$. Gray shaded areas are representing regions where $f(d)>F(d)$ and
the criterion for producibility is fulfilled.}
\label{force-energy}
\end{figure}

Next, for an Au chain in the thermodynamic limit ($N \rightarrow \infty$) we introduce some important 
general concepts necessary for a detailed analysis of the results. As it is apparent from
Fig.~\ref{force-energy} and eq.~(\ref{P2}) two distinct regions of producibility can arise:
one in the domain of zigzag chains (angle $\alpha > 0$), and the second in the region of
straight chains (angle $\alpha \approx 0$). Accordingly, in both regimes the maximal sustainable
force $F_0$ is observed at two different values of the inflection point $\hat{d}$. And while both
parameters undergo large changes depending on the system, $F_0$ and $\hat{d}$
of the straight chains have a clear physical meaning and can provide important understanding of the
most general trends of the chain formation. Hence $-$ if not specified explicitly $-$ $F_0$ and
$\hat{d}$ will refer to the straight chain domain from here on.

\begin{figure}[t!]
\begin{center}
\includegraphics[width=8.5cm]{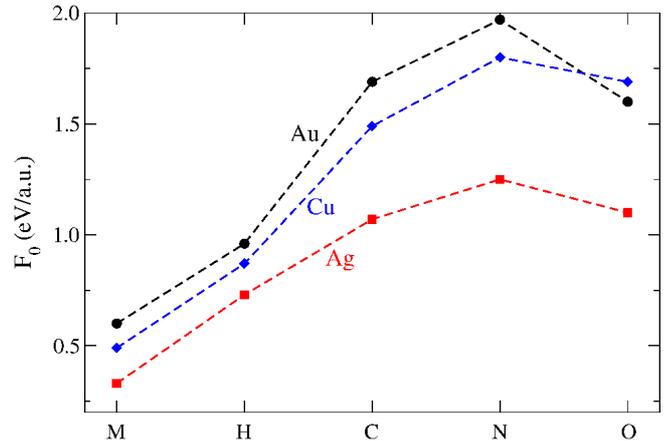}
\end{center}
\caption{(Color online). Break force $F_0$ for Au, Ag, and Cu pure
and impurity-assisted chains. Along  the $x$-axis "M" labels
the pure chain, while "H", "C", "N", and "O" $-$ chains with corresponding impurities.}
\label{breaking-force}
\end{figure}

The most obvious influence of considered impurities on the elongation process can be seen in the
changes in the break force $F_0$ when compared to the case of the pure chains.
From Fig.~\ref{breaking-force}, in which the break force for each of the studied cases is given,
we first of all observe that in pure noble-metal chains a stronger binding and correspondingly larger
values of $F_0$ occur for Au and Cu, while Ag chains have the weakest binding. This is a direct
consequence of shallower energy profiles for Ag chains, discussed in section~\ref{single-atom}.
The fact that Au chains exhibit the largest value of $F_0$ among all NMs is in direct correspondence
to experiments.~\cite{Ruitenbeek01,Alex08} 
When environmental impurities are added to the system, the bonds in all cases become stronger 
and the break forces larger owing to the pronounced covalent bonding in NM-X chains, as compared
to metallic binding in pure chains. Noticeably, among all impurities H atoms cause smallest
strengthening of the bonds, considering an absence of directionality of $s$ orbitals, in 
contrast to the case of $p$ impurities. On the other hand, adding $p$ impurities to the NM chains
leads to an approximately three-fold increase in the break force. Here, the most significant
influence on $F_0$ can be seen for NM chains with nitrogen impurities.

\begin{figure*}
\begin{center}
\begin{tabular}{ccccc}
\includegraphics[width=0.32\textwidth]{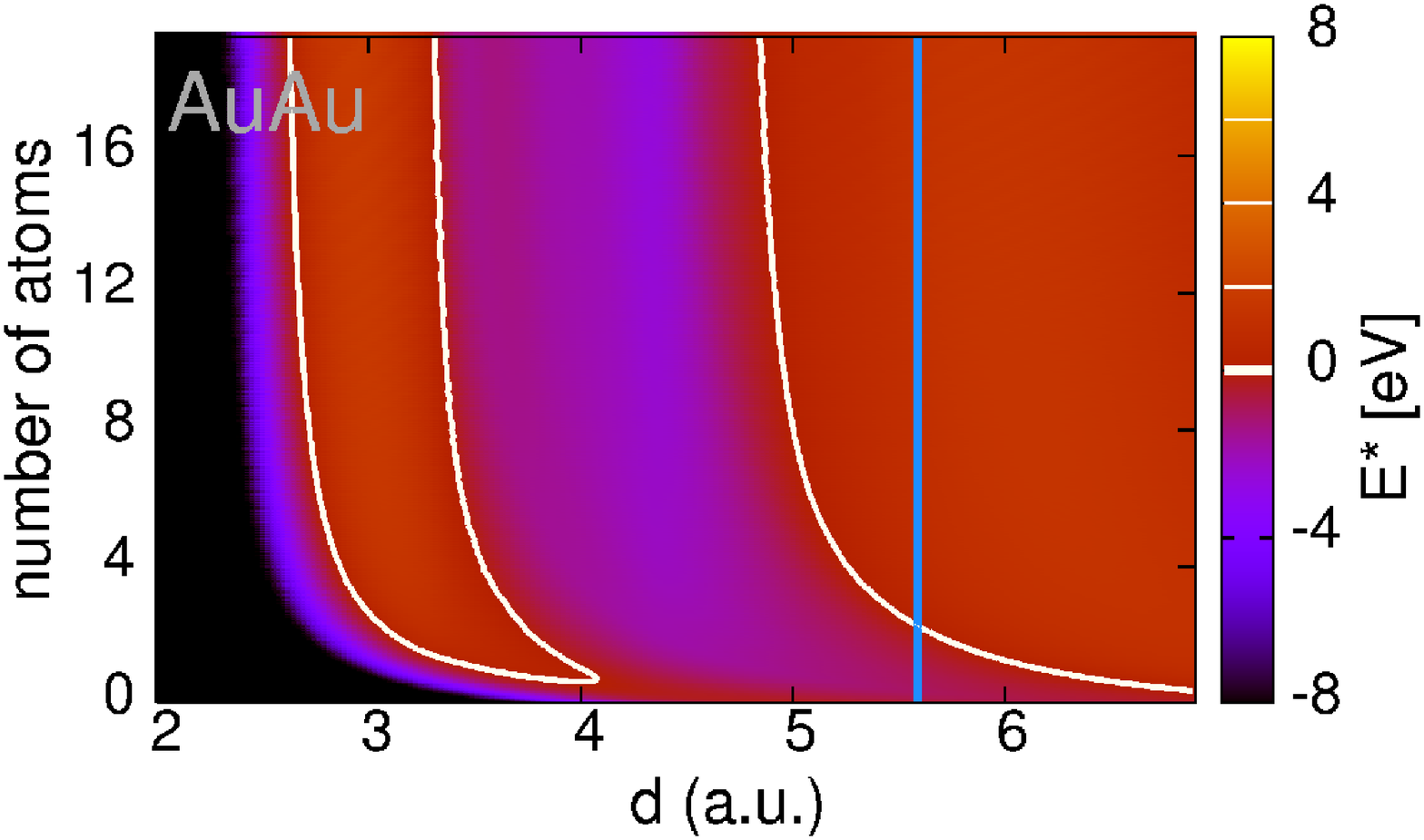} & & \includegraphics[width=0.32\textwidth]{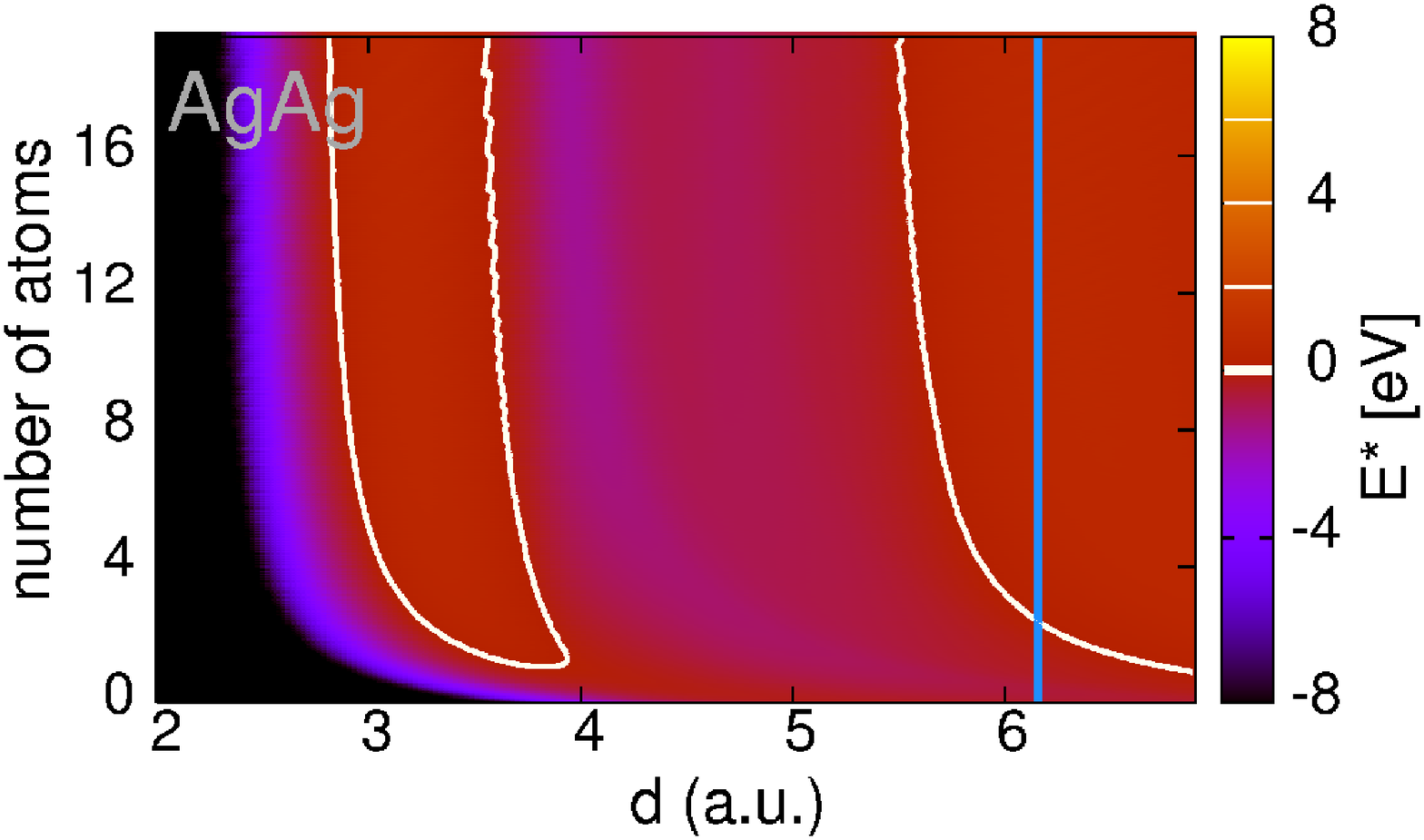} & & \includegraphics[width=0.32\textwidth]{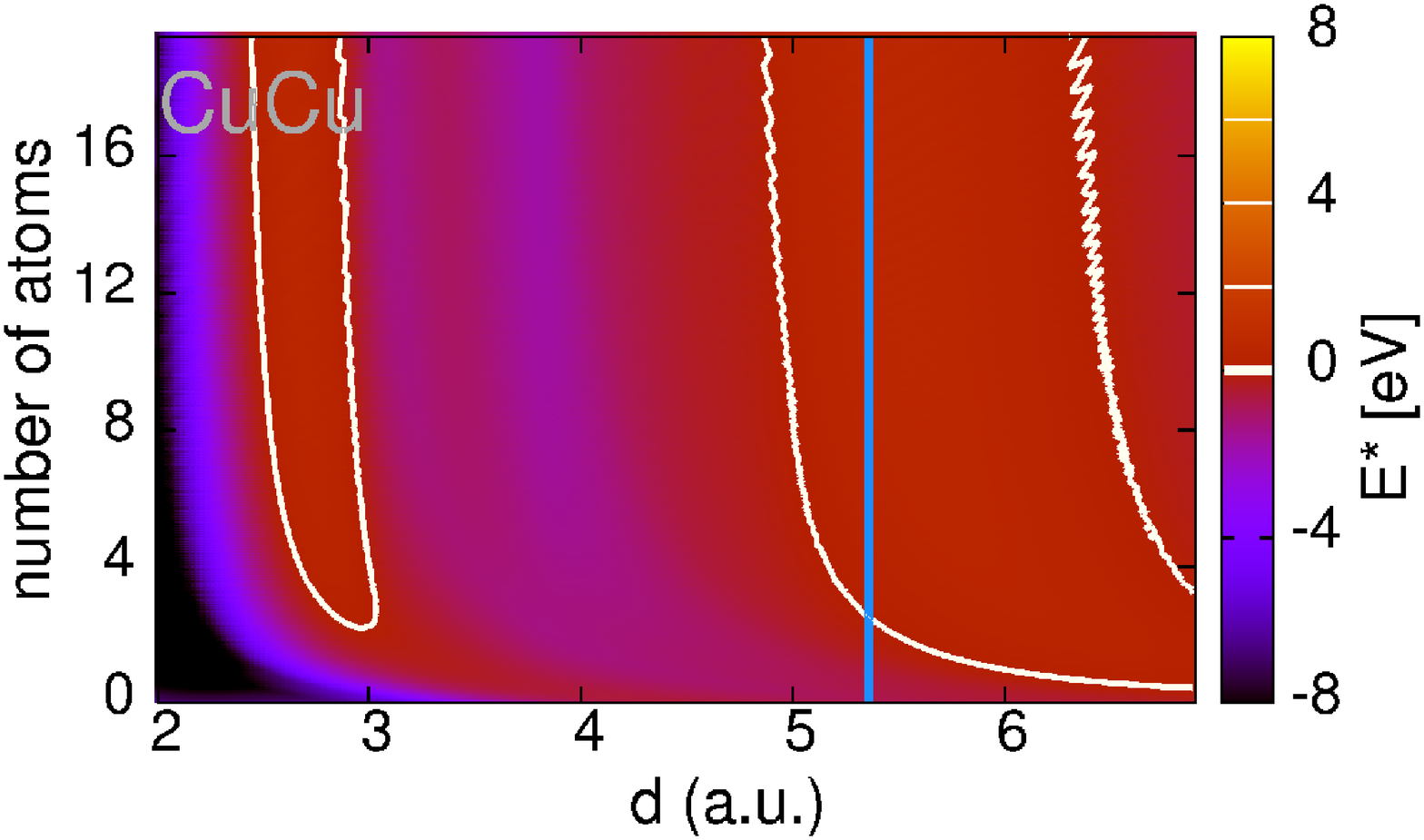} \\
\includegraphics[width=0.32\textwidth]{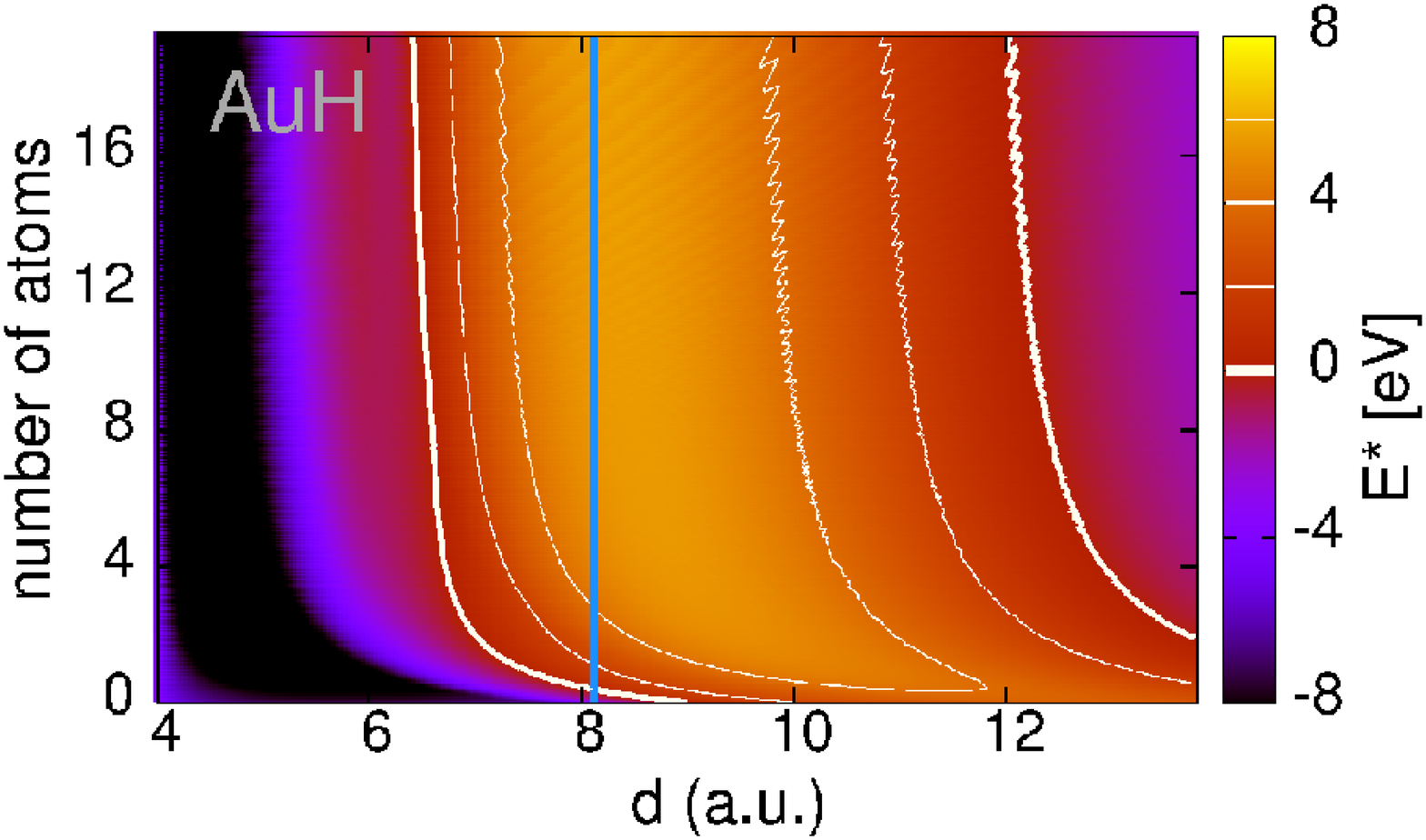}  & & \includegraphics[width=0.32\textwidth]{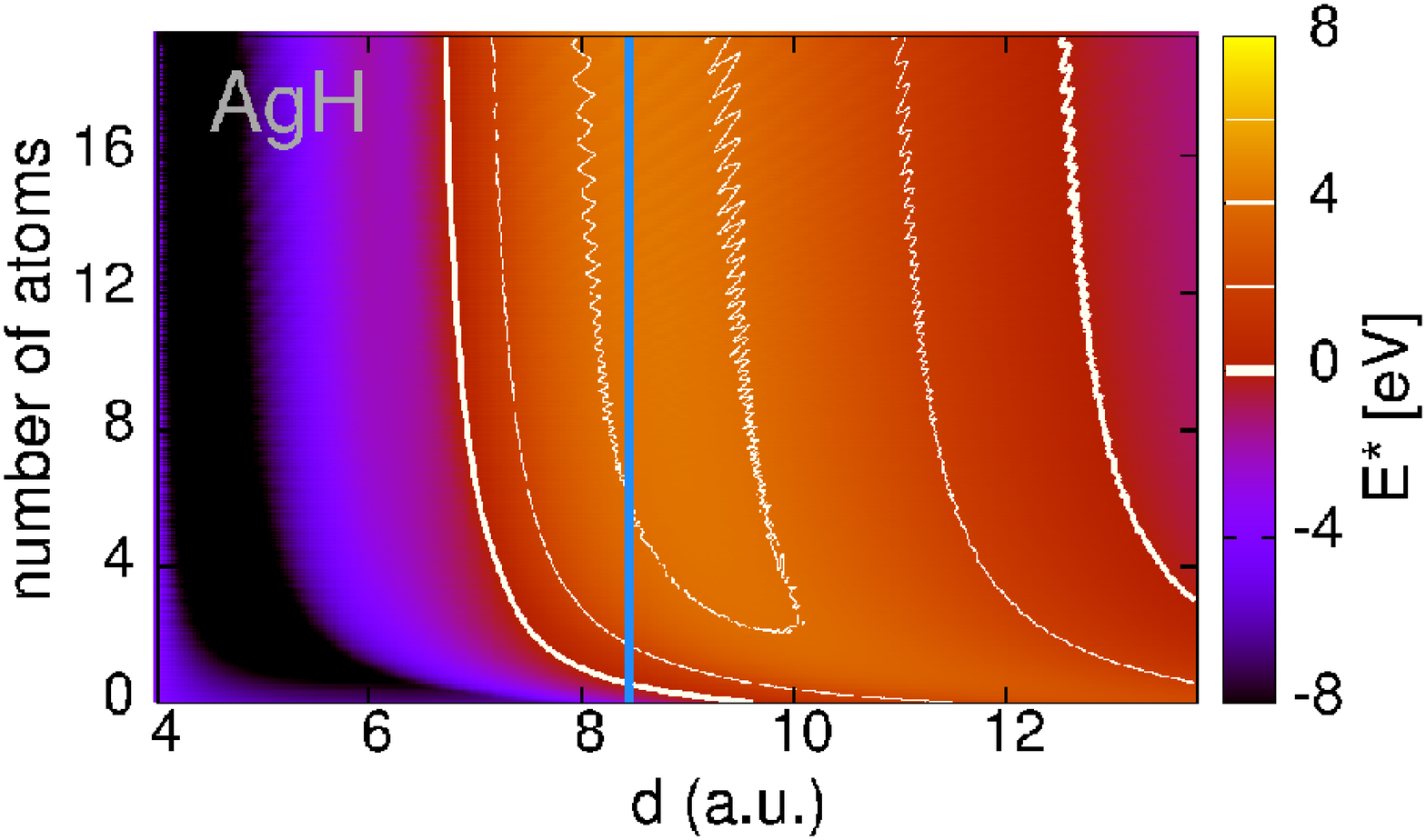}  & & \includegraphics[width=0.32\textwidth]{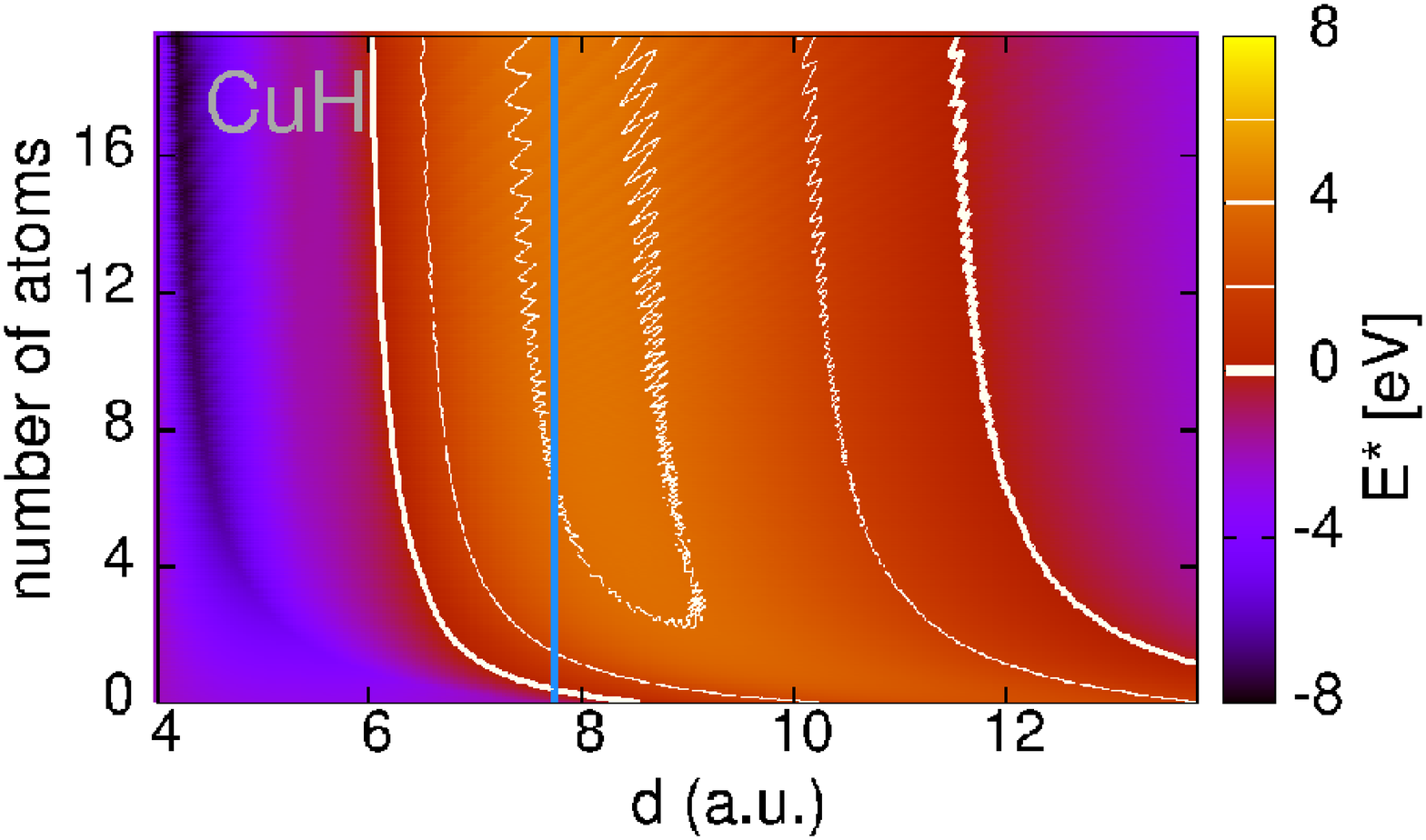}  \\
\includegraphics[width=0.32\textwidth]{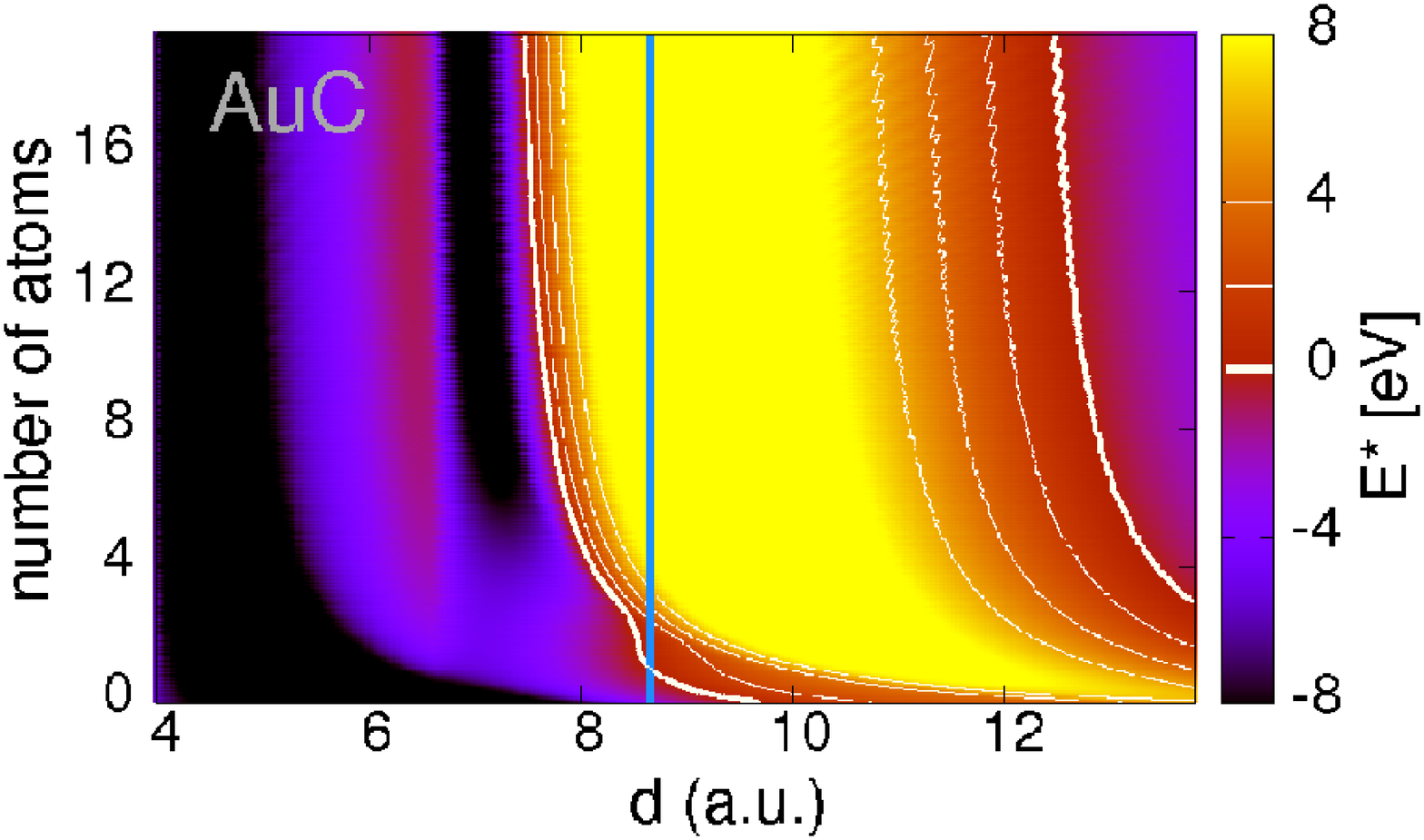}  & & \includegraphics[width=0.32\textwidth]{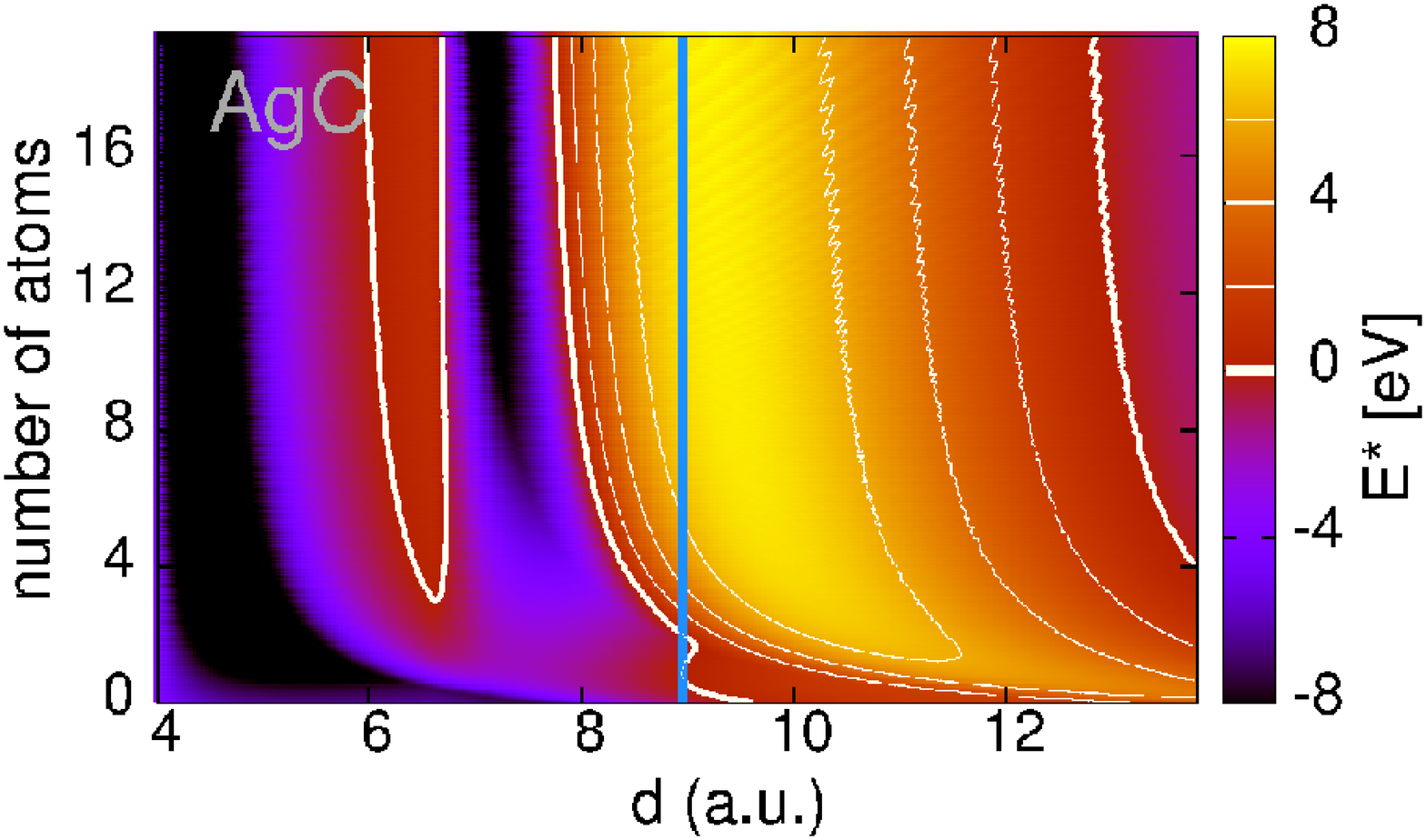}  & & \includegraphics[width=0.32\textwidth]{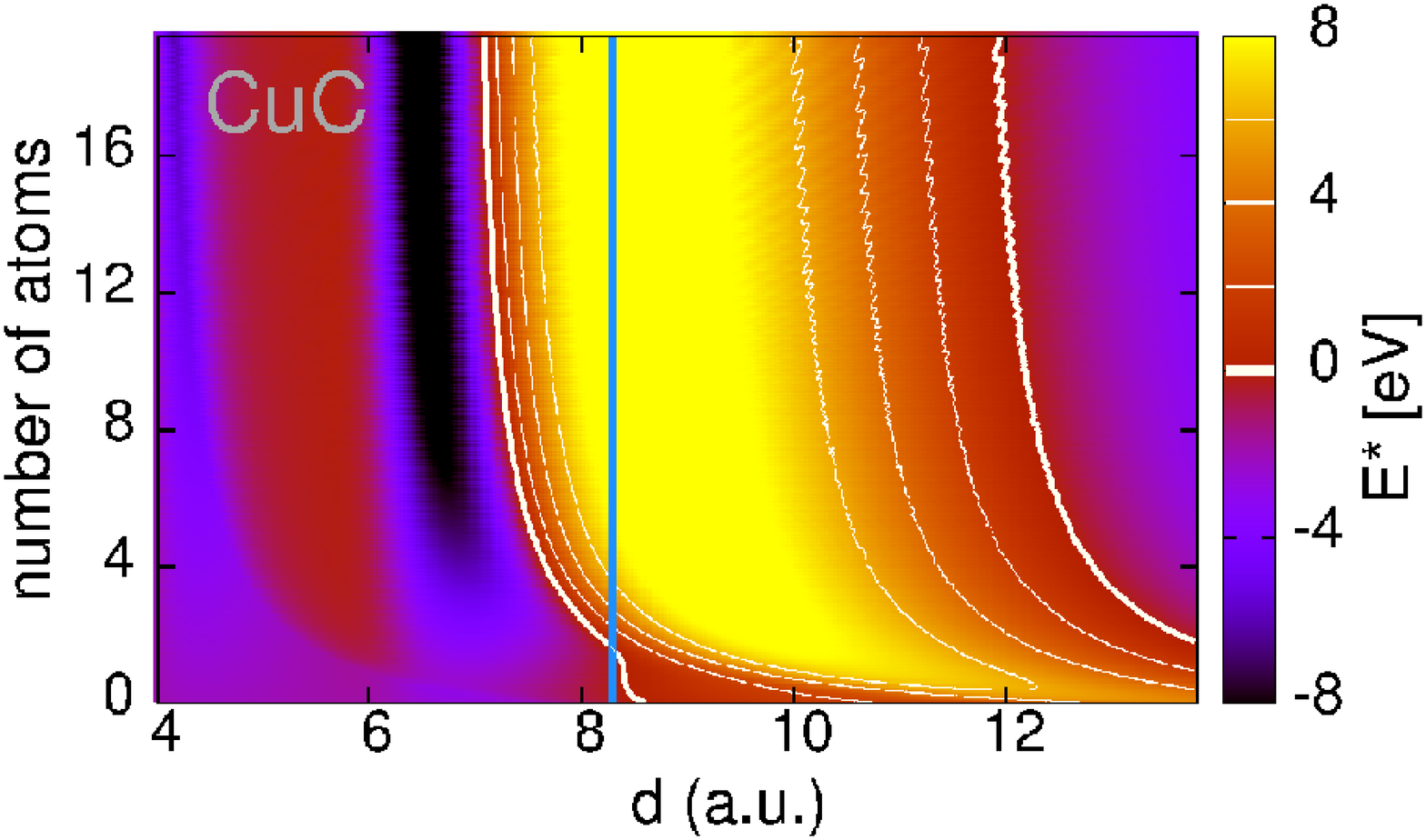}  \\
\includegraphics[width=0.32\textwidth]{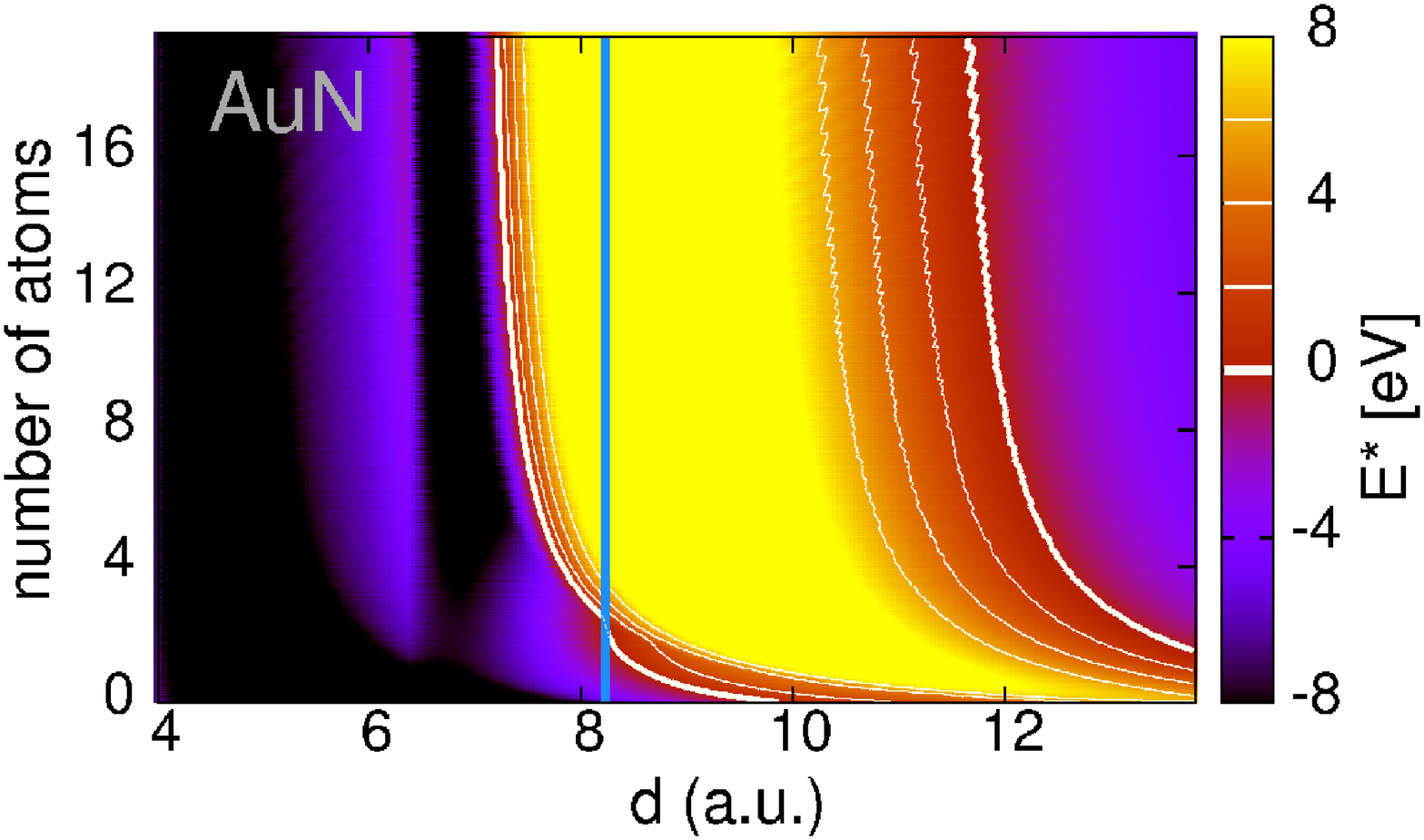}  & & \includegraphics[width=0.32\textwidth]{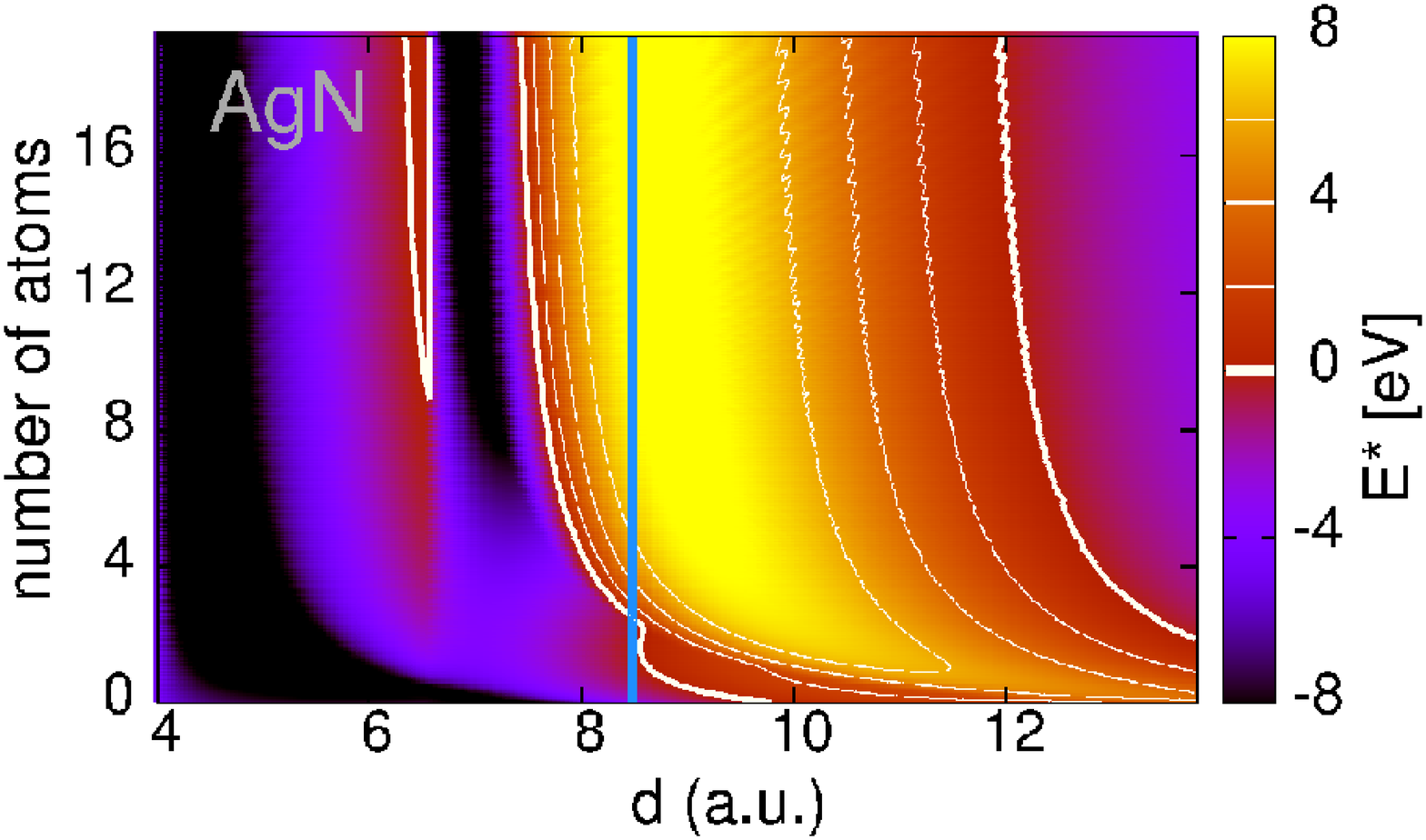}  & & \includegraphics[width=0.32\textwidth]{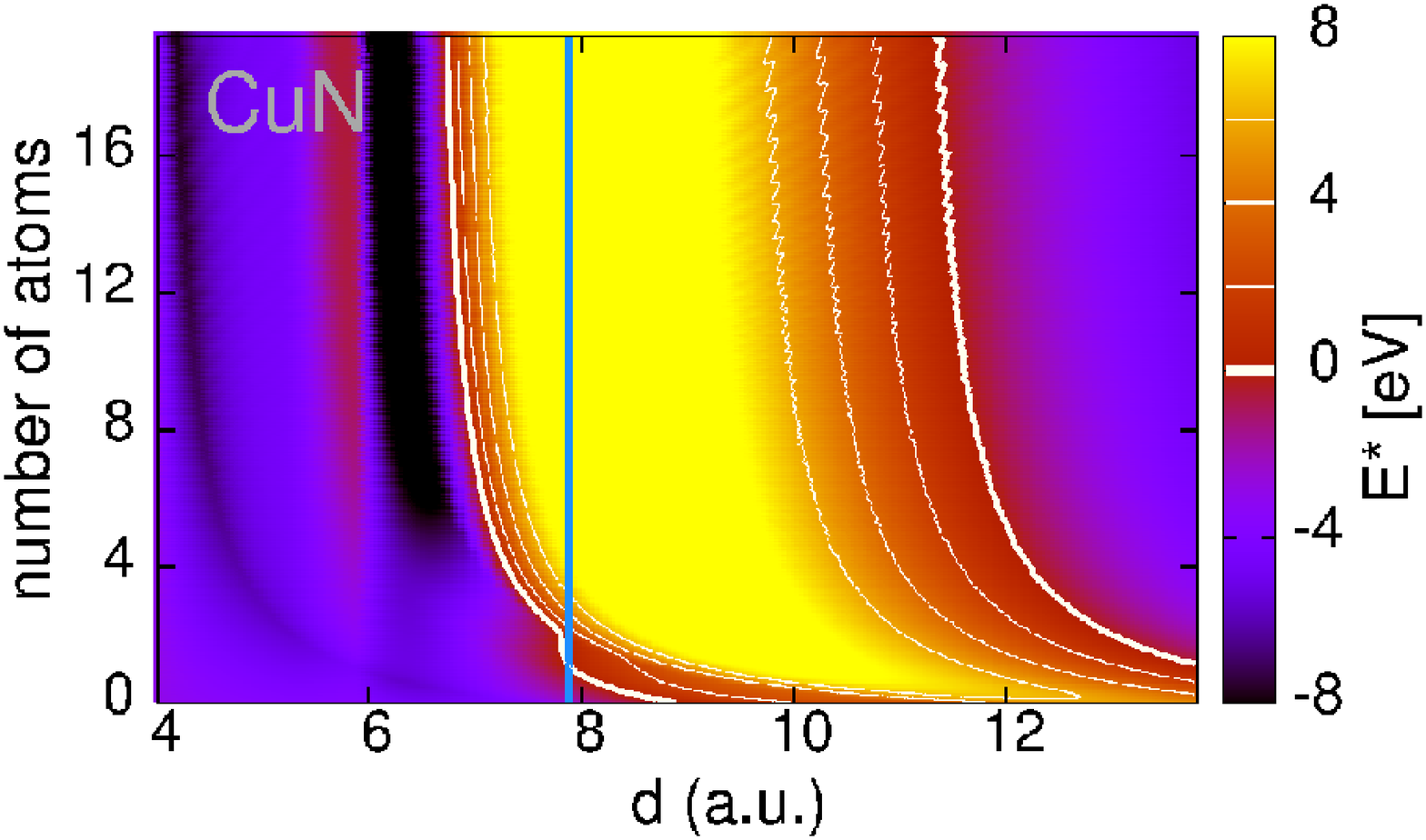}  \\
\includegraphics[width=0.32\textwidth]{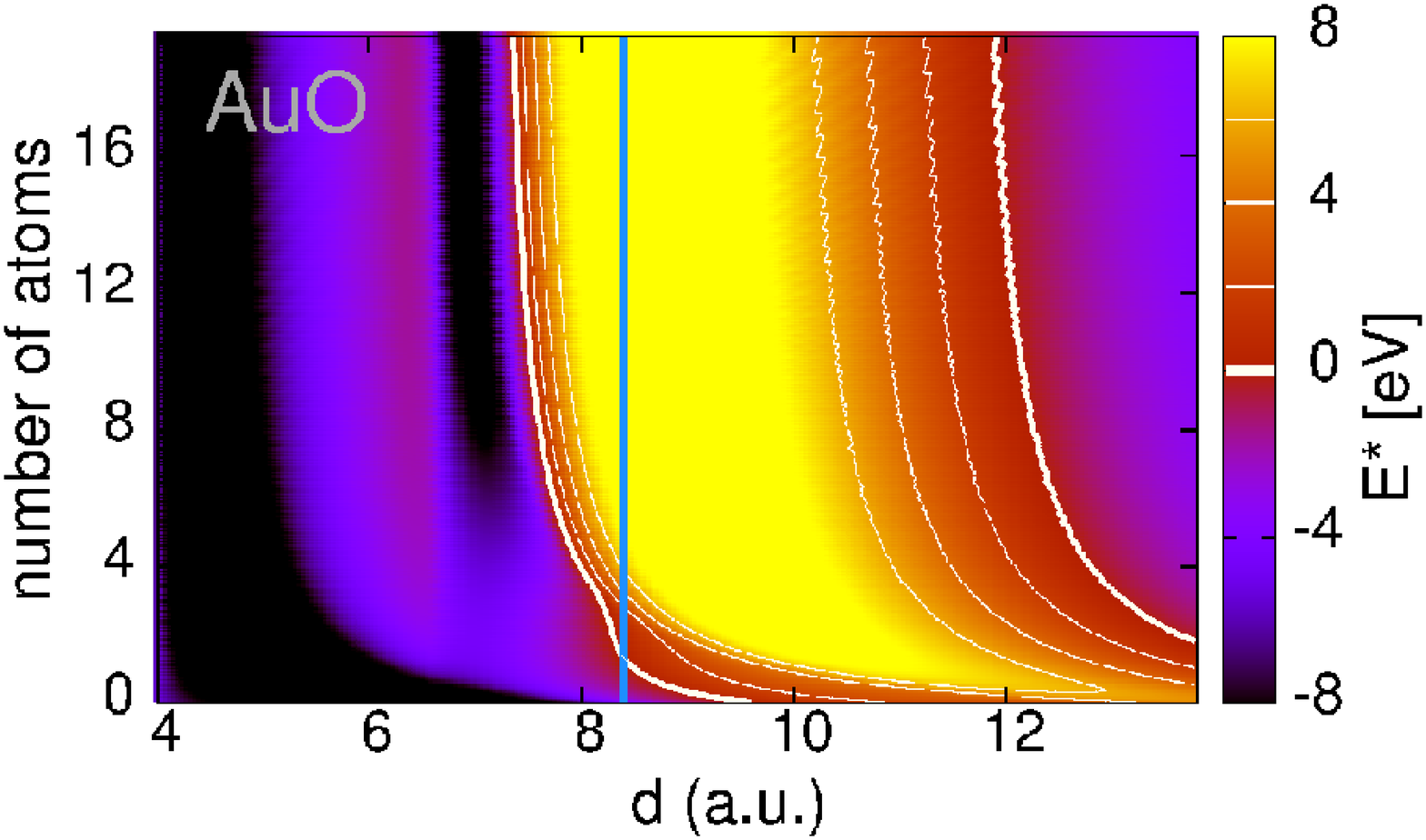}  & & \includegraphics[width=0.32\textwidth]{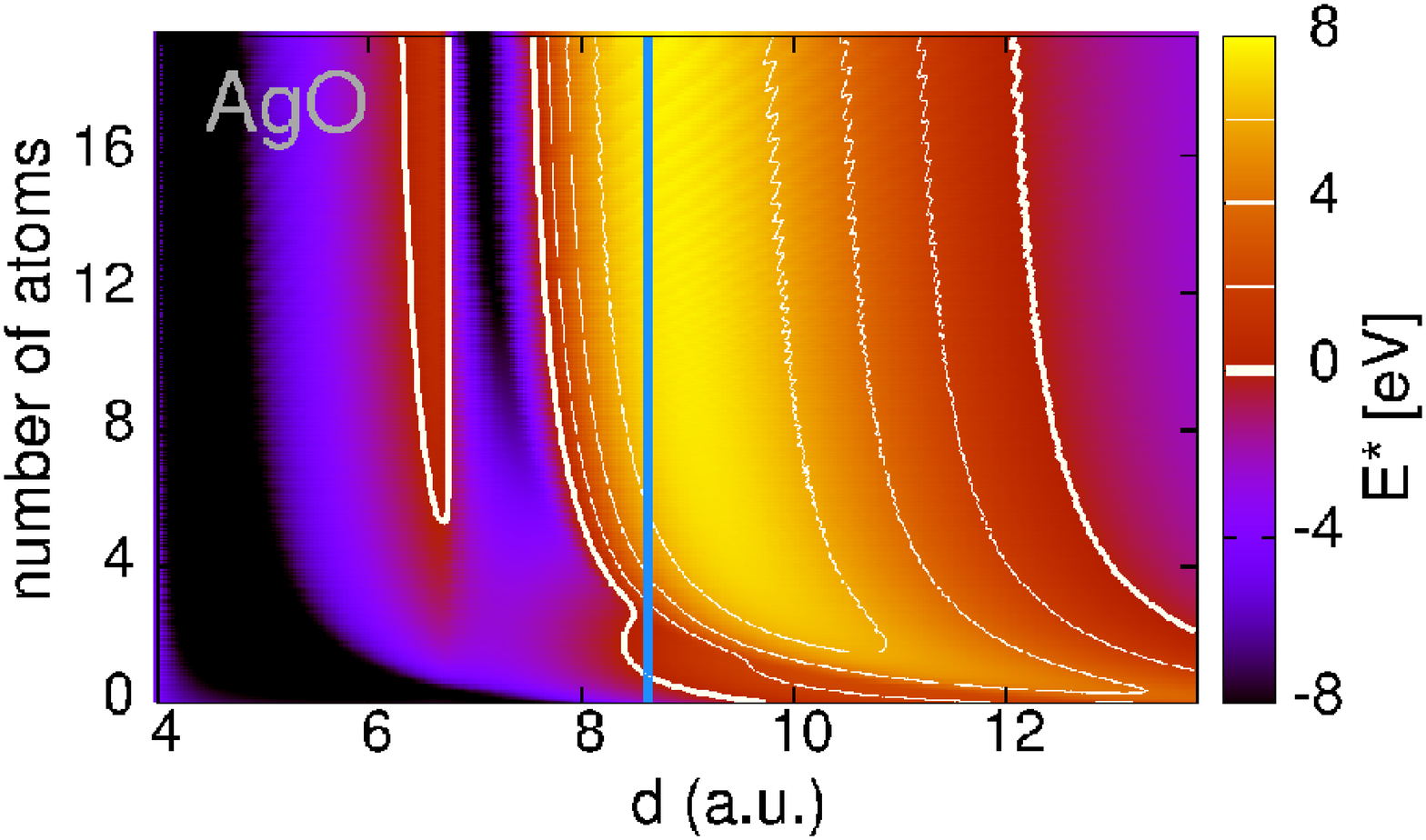}  & & \includegraphics[width=0.32\textwidth]{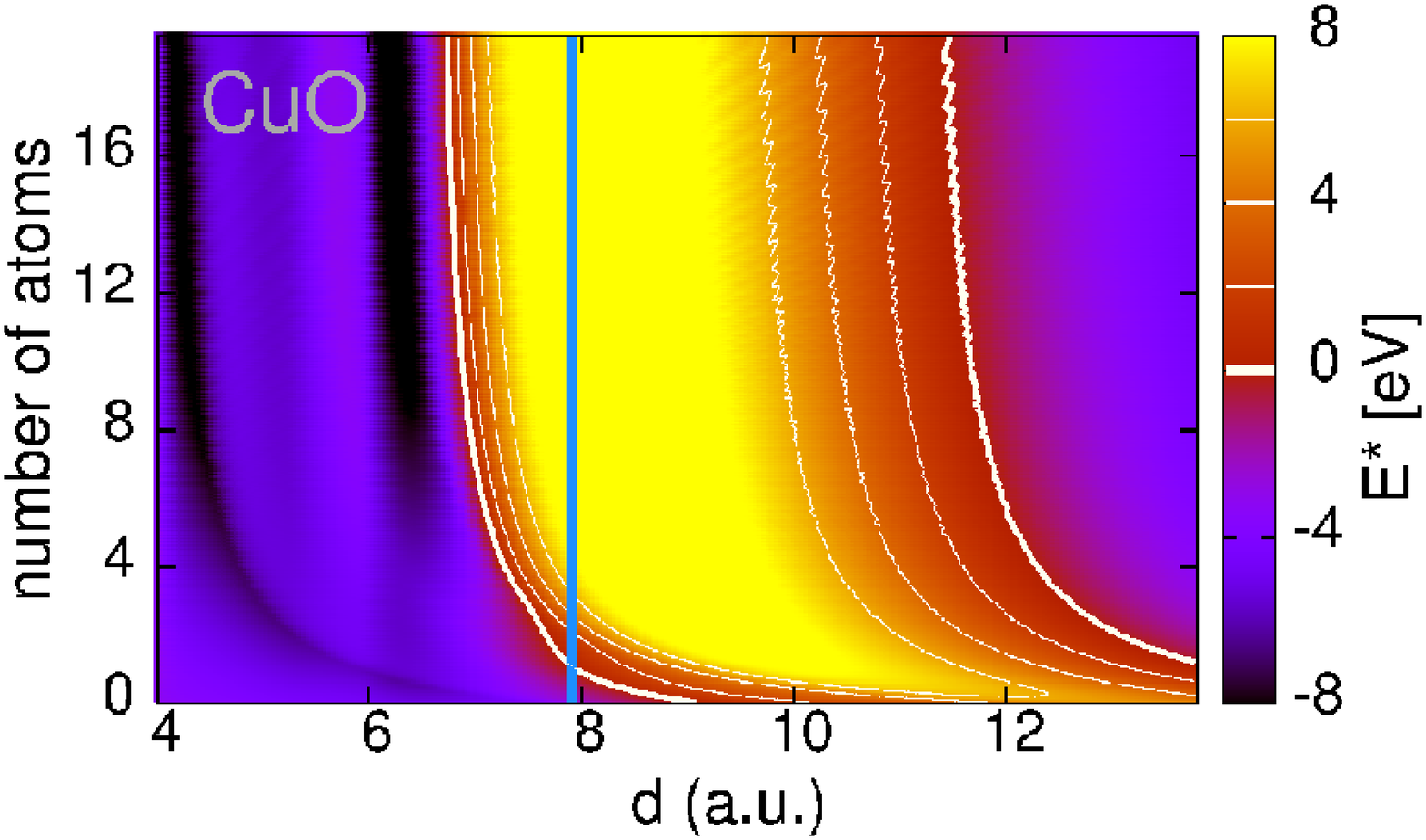}
\end{tabular}
\end{center}
\caption{Energy landscape $E^*(d,N)$ as a function of interatomic distance ($d$) and 
         number of atoms in the chain ($N$) describing the chain producibility for pure and 
         impurity-assisted noble metal chains, where $E^*$ is defined by the criterion for producibility,
         eq.~(\ref{E*}).
         From left to right Au, Ag and Cu chains are displayed, from top to bottom, pure wires and
         H, C, N and O assisted chains.
         Thick white lines mark the $E^{*}=0$ isolines indicating regions of producibility ($E^*(d,N) \textgreater 0$). 
         Additional thin isolines are drawn at $E^*(d,N)=2$~eV, 4~eV, 6~eV, and 8~eV in each panel. 
         Note that regions which are outside of the interval [-8~eV, 8~eV] are shown in the same 
         color-coding as the minimal and maximal value of this interval. 
         To rate the stability of chains, for each panel the inflection point $\hat{d}$ 
         in the linear regime of the binding energy potential is depicted by thick blue vertical lines.}
\label{Au-P}
\end{figure*}

Next, we perform a detailed analysis of the criterion of producibility for the NM and NM-X 
chains, as
given by eq.~(\ref{E*}). In Fig.~\ref{Au-P} we plot the energy $E^{*}$ as a function of the
interatomic distance in the chain, $d$, and the number of chain atoms, $N$, both for pure
NM chains, as well as NM chains with H, O, C and N impurities. The region of the $(N,d)$
phase space where $E^{*}>0$ provides the parameters of the chain configuration for which
the elongation by one atom, or one NM-X pair of atoms, is possible. Additionally, the absolule
value of the $E^{*}$ energy gives an estimate of how probable or how improbable the event of
elongation is. Correspondingly, in Fig.~\ref{Au-P} the absolute value of $E^{*}$ is given
with appropriate color coding. The ability to estimate the $E^{*}$ value is extremely useful,
since in experiments the energy barrier for an atom to enter the chain can fluctuate, depending
on the geometry of the contacts. Therefore, high positive values of $E^{*}$ provide higher
credibility to the occurrence of a single event of chain elongation at a particular point
in the phase diagram for which the P-criterion is satisfied.

An important aspect of the chain creation process to consider is the stability of 
the suspended chains. The chain's stability in a break junction is a very complicated 
issue, since here the primary importance is due to the interplay of local tip geometry,
phonon spectra of the finite chains, as well as temperature and applied bias.
To model the chain's breaking process is therefore a highly non-trivial task, which
is specific to each attempt of chain formation, and which cannot be undertaken in the
present study. Here, we restrict ourselves to a simple but rather transparent way
of quantifying the stability of the chains by referring to the value of their inflection 
point, $\hat{d}$,~i.e.~the interatomic distance at which the maximal force $F_0$ is
sustained. The value of $\hat{d}$, which can be interpreted as the 
minimal interatomic distance at which the chain will be broken for a given $N$, 
is presented for each of the cases by a thick vertical line in the phase diagram of Fig.~\ref{Au-P}. 

As we can see for pure NM chains in Fig.~\ref{Au-P}, the position of the boundary of 
the P-region with respect to $\hat{d}$ is very similar for Cu, Ag and Au. On the
other hand, for NM-X chains, the value of $\hat{d}$ varies in a comparatively narrow 
interval of distance between 7.9 and 8.9 bohr, Fig.~\ref{Au-P}. This means that
the properties of considered chains with respect to their stability are comparatively
constant. This observation is in accordance to a previous finding of ref.~\cite{Alex08}
in which the stability of linear suspended $5d$ transition metal chains has been
analyzed within a better approximation.

Let us first take a closer look at the phase diagrams for pure NM chains in Fig.~\ref{Au-P}.
First we observe that all chains exhibit large areas of producibility, both for linear and
zigzag geometries. This is a striking new finding, which implies that in the general case 
of even more complex geometries of suspended wires more than one structurally stable 
phases favoring chain elongation can co-exists. This suggests that in experiment the
co-existence of such phases can be used to enhance the elongation properties of the chains 
by switching from one phase to another via rapid pulling or pushing the electrodes apart.
This technique can be particularly efficient, given that the stability of the phases
and the respective elongation probabilities (proportional to $E^{*}$) can depend 
strongly on the details of the particular elongation attempt.

Secondly, we observe that chains consisting of more than 10 atoms can be described very
well in the thermodynamic limit. Such length scale is rarely reached in experiment,\cite{Kizuka08}
however, which underlines the importance of analyzing chain creation trends as a function of $N$. 
On this side, the description of the chain formation process 
at its initiation with one to three atoms in the chain is clearly improved by extending
the CF model~\cite{Alex08} to zigzag geometries: The strict assumption
of linear bonds leads to an additional cost in energy for relaxations below the 
equilibrium interatomic distance. In our extended model, the chains relax at such short 
interatomic distances into a zigzag geometry with a comparable or even favorable energy 
contribution. Consequently, for Ag and Au the criterion for producibility can be fulfilled 
for smaller $N$. Generally speaking, this underlines the importance of structural versatility
in a break junction experiment for chain elongation.
Third, both linear and zigzag phases of producibility have largest values of $E^*$ energy for 
Au. The corresponding areas become less extended and exhibit somewhat smaller values of $E^*$ 
when going from Ag to Cu. This trend can be direclty related to the values $F_0$ in 
 Fig.~\ref{breaking-force} and its interplay with $\Delta E_{\rm lead}$.

Going now to NM chains with H impurities (second panel in Fig.~\ref{Au-P}), 
we observe wide regions of producibility only in the 
linear regime for all metals, in accordance to small ground state zigzag angles in this case. 
And while the width of the P-region is roughly the same for Cu,
Ag and Au, the probability of chain elongation, characterized by the energy $E^{*}$, 
is noticeably larger for Au, than for Ag and Cu. In any case, the $E^{*}$ values, reaching 
as much as 5~eV for Au, are significantly larger than corresponding values for pure chains.
Additionally, for all $N$ the area of producibility can be entered upon 
increasing the interatomic distance in the chain without exceeding the inflection point.  
Overall, this leads to the prediction of a strongly enhanced tendency towards chain formation 
and elongation in NM break juncction with H in the atmosphere.

Finally, we turn to the case of NM chains with $p$ impurities C, O and N (three lower panels
of Fig.~\ref{Au-P}). As we analyzed previously, adding $p$ impurities to the NM chains results
in a three-fold increase in the values of the break force (Fig.~\ref{breaking-force}). This
results in very wide regions of producibility for all chains with very high values of 
$E^{*}$ reaching over the threshold of 8~eV. These values are significantly larger than
those for pure chains and even for chains with H adatoms. While the width of the P-region
in the linear regime is similar for all NMs, the values of $E^{*}$ are somewhat smaller for
Ag, than for Cu and Au, which can be attributed to the different break forces for these elements,
see Fig.~\ref{breaking-force}. Characteristically, for all cases the part of the P-region in the
linear regime for which the interatomic distance lies below the inflection point, is significant,
thus ensuring stucturally stable elongation.

As apparent from Fig.~\ref{X-NM}, contrary to the case of Cu and Au, Ag chains with C, N and O
impurities exhibit two pronounced energy minimuma for the zigzag and linear regimes, separated by
a local energy maximum. In terms of the P-criterion this results in finite zigzag producibility
regions for Ag, in addition to the P-region in the linear regime. And although the values of $E^{*}$
are modest for zigzag chains, the area in the phase space where the zigzag chains can be elongated
is clearly visible especially for C- and O-assisted Ag chains. As discussed previously for pure chains,
the second region of producibility can open the way to more efficient elongation. Despite the fact that
shorter Ag chains can be grown in the linear regime only, for larger $N$ rapid pushing of the tips 
towards each other might lead to a jump from the linear into the zigzag mode with further chain 
growth in the latter regime of higher stability. Clear indication of the zigzag-like areas can
be also seen for smaller $d$ in AuC, AuO, CuN, and in particular CuC cases, although the energy 
$E^{*}$ is negative and reaches as much as $-$1~eV in these regions. Nevertheless, depending on
the details of the chain growth and geometry of the tips, we believe that such regions can acquire
positive energy $E^{*}$ and lead to longer chains in analogy to break junctions of Ag. 
Overall, we conclude that the tendency to elongation of NM chains with $p$ impurities is
enhanced when compared to pure NM chains and NM chains with hydrogen. 

\section{Conclusions}
\label{conclusions}

We have presented the generalization of the model for formation of
zigzag and impurity assisted chains in break junction experiments. 
On the basis of the extended model we investigated the growth of noble metal
wires with H, O, C and N impurities, taking the input from first-principles calculations.
Our most important observation is that impurity assisted chain growth leads 
to a strongly enhanced tendency towards chain elongation, when compared to pure noble metal chains.
Moreover, we reveal a distinct difference between the influence of the $s$ and $p$ type of impurities
on chain creation, finding that $p$ impurities lead to a higher chain elongation probability owing
to a strong directional bonding. Importantly, we predict that the existence of more than one stable
phases favoring chain elongation can occur in some cases, and suggest to employ this phenomenon
for production of longer chains in experiment. By utilizing one of the most important advantages of 
our model we are able to translate our findings into phase diagrams for successful chain creation 
explicitly providing the values for the interatomic distances and the length of the chains for which
the elongation can be achieved.

\acknowledgments{
The authors kindly thank Prof. J.~M.~van Ruitenbek and Prof. E.~Scheer for inspiring and fruitful discussions.
Y.~M.~gratefully acknowledges funding under the HGF-YIG Programme VH-NG-513 and S.~D.~N acknowledges funding from
Conicet, PIP00258.} 

\bibliography{chains}

%\bibliographystyle{apsrev}
%\bibliographystyle{apsrev4}
%\bibliography{chains}

\end{document}